\documentclass[aps,prd,twocolumn,showpacs,superscriptaddress,groupedaddress,preprintnumbers,nofootinbib,10pt]{revtex4-1}
\usepackage{graphicx}  
\usepackage{dcolumn}   
\usepackage{bm}        
\usepackage{amssymb}   
\usepackage[latin1]{inputenc}
\usepackage{amsmath}
\usepackage{amsfonts}
\usepackage{amssymb}
\usepackage{setspace}
\usepackage{mathtools}
\usepackage{pstricks}
\usepackage{color}
\usepackage{hyperref}

\providecommand{\f}[2]{\frac{{#1}}{{#2}}}
\usepackage{accents}
\newcommand{\ubar}[1]{\underaccent{\bar}{#1}}
\newcommand{\ee}[1]{\begin{equation}#1\end{equation}}
\newcommand{\ea}[1]{\begin{align}#1\end{align}}

\def\undertilde#1{\mathord{\vtop{\ialign{##\crcr
$\hfil\displaystyle{#1}\hfil$\crcr\noalign{\kern1.5pt\nointerlineskip}
$\hfil\tilde{}\hfil$\crcr\noalign{\kern1.5pt}}}}}

\begin{document}
 
\preprint{KCL-PH-TH/2017-14}
\title{De Sitter Stability and Coarse Graining}


\author{T.~Markkanen}\email{t.markkanen@imperial.ac.uk}
\affiliation{Department of Physics, Imperial College London, SW7 2AZ, UK}
\affiliation{Department of Physics, King's College London, Strand, London WC2R 2LS, UK}
\date{\today}

\begin{abstract}
We present a 4-dimensional back reaction analysis of de Sitter space for a conformally coupled scalar field in the presence of vacuum energy initialized in the Bunch-Davies vacuum. In contrast to the usual semi-classical prescription, as the source term in the Friedmann equations we use expectation values where the unobservable information hidden by the cosmological event horizon has been neglected i.e. coarse grained over. It is shown that in this approach the energy-momentum is precisely thermal with constant temperature despite the dilution from the expansion of space due to a flux of energy radiated from the horizon. This leads to a self-consistent solution for the Hubble rate, which is gradually evolving and at late times deviates significantly from de Sitter. Our results hence imply de Sitter space to be unstable in this prescription. The solution also suggests dynamical vacuum energy: the continuous flux of energy is balanced by the generation of negative vacuum energy, which accumulatively decreases the overall contribution. Finally, we show that our results admit a thermodynamic interpretation which provides a simple alternate derivation of the mechanism. For very long times the solutions coincide with flat space. \end{abstract}
\maketitle


\section{Introduction}
The de Sitter spacetime is one of the most analytically tractable examples of a genuinely curved solution to Einstein's field equation. De Sitter space is not only of academic interest since in the current cosmological context the exponentially expanding de Sitter patch is believed to describe the evolution of the Universe soon after the Big Bang during cosmological inflation and at very late times when Dark Energy has begun to dominate over all other forms of energy.

The potential instability of de Sitter space in quantized theories has been investigated in a variety of different approaches and models over a span of more than 30 years \cite{Tsamis:1996qq2,Tsamis:1996qq3,Geshnizjani:2002wp,Geshnizjani:2003cn,Greenwood:2010mr,Tsamis:1996qq, Tsamis:1996qq0,Tsamis:1996qq1,Brandenberger:1999su,ford,Polyakov:2007mm, Polyakov:2007mm2,akh,Marolf:2010zp,Marolf:2010nz,ArkaniHamed:2007ky, clif,Antoniadis:1985pj,Mottola:1984ar,Mottola:1985qt, Anderson:2013ila,Anderson:2013ila1,Kachru:2003aw, Rajaraman:2016nvv,Goheer:2002vf,Padmanabhan:2002ji,Finelli:2011cw,Firouzjaee:2015bqa,Brown:1988kg,Bousso:2000xa, Markkanen:2016vrp,Markkanen:2016aes,Dvali:2017eba,Xue:2014kna,Bavarsad:2016cxh}, recently in \cite{Markkanen:2016jhg} where we refer the reader for more references. To the best of our knowledge, at the moment the issue still lies unresolved. If de Sitter space were unstable to quantum corrections and could indeed decay, this could provide an important mechanism for alleviating the cosmological constant problem and perhaps also the fine-tuning issues encountered in the extremely flat inflationary potentials that are required by observations. Most definitely, a de Sitter instability would have a profound impact on the fate of the Universe since it rules out the possibility of an eternally exponentially expanding de Sitter space as classically implied by the $\Lambda$CDM concordance model.

One of the main motivations behind the original calculation for the evaporation of black holes in \cite{Hawking:1974sw,Hawking:1974rv} was the discovery of their thermodynamic characteristics \cite{Bekenstein:1973ur,Bekenstein:1974ax}, in particular the connection between the black hole horizon and entropy: the fact that black holes evaporate implies that they can also be ascribed temperature and understood as thermodynamic objects. Like a black hole de Sitter space also possesses a horizon beyond which a local observer cannot see, which was famously in \cite{Gibbons:1977mu} shown to lead to a thermodynamic description of de Sitter space analogously to a black hole. Currently, the thermodynamics of spacetime horizons is established as a mature, well-studied subject \cite{Padmanabhan:2003gd,Padmanabhan:2002ha,Sekiwa:2006qj,Padmanabhan:2002ji, Bousso:2001mw,Jacobson:1995ab,Cai:2005ra,Padmanabhan:2002sha}. Based on thermodynamic arguments the seminal study \cite{Gibbons:1977mu} concluded that unlike black holes de Sitter space is stable. However quite interestingly, also by invoking thermodynamic concepts in the equally impactful work \cite{Padmanabhan:2002ji} it was argued that the de Sitter horizon in fact does evaporate.\footnote{The argument can be found in section 10.4 of \cite{Padmanabhan:2002ji}.} 

As in the original black hole evaporation calculation \cite{Hawking:1974sw} we make use of semi-classical gravity -- often referred to as quantum field theory in curved spacetime \cite{Birrell:1982ix,Parker:2009uva} -- in order to provide a first principle calculation of the stability of de Sitter space. 
Our approach allows one to study how the quantized matter back reacts on the classical metric by using the semi-classical versions of Einstein's equation. In situations where the quantum nature of gravity is subdominant this is expected to give reliable results. Specifically we will focus on the cosmologically most relevant coordinate system, the Friedmann--Lema\^{i}tre--Robertson--Walker (FLRW) line element describing an expanding, homogeneous and isotropic spacetime. This line element results in the Friedmann equations allowing a straightforward analysis of back reaction, studied for example in \cite{Herranen:2016xsy,Herranen:2015aja,Marozzi:2012tp, Koivisto:2010pj,Rigopoulos:2013exa,Rigopoulos:2016oko, Bilandzic:2007nb,Markkanen:2012rh,Herranen:2013raa,Markkanen:2016aes}. The FLRW coordinates are rarely included in discussions of the thermodynamics of horizons, although see \cite{Cai:2005ra,Frolov:2002va,Singh:2013pxf}. 

The decoherence program asserts that the ubiquitous disappearance of macroscopic quantum effects -- commonly known as the quantum-to-classical transition -- stems from the observationally inaccessible environmental sector that in any realistic set-up is always present \cite{Joos:1984uk, Zurek:1982ii,Zurek:1981xq,Zeh:1970zz,Kiefer:1998qe,Kiefer:2001wn}. Using this mechanism as a motivation recently in \cite{Markkanen:2016jhg} a modification to the usual prescription for semi-classical gravity was explored where in the Einstein equation one implements coarse grained expectation values calculated by including only those states that are observable. It was shown that if a part of the density matrix may be characterized as unobservable and is neglected from the quantum averaging this generically leads a qualitatively different behaviour for the expectation value for the energy-momentum in de Sitter space compared to the usual approach: it implies non-trivial back reaction with an evolving Hubble rate, even when as the initial condition one uses the manifestly de Sitter invariant Bunch-Davies vacuum. The procedure of tracing over unobservable states, in addition to decoherence studies, is often implemented in calculations involving spacetimes with horizons such as black holes and Rindler space \cite{Unruh:1976db,Unruh:1983ms,Crispino:2007eb} and is a key element of the information paradox \cite{Hawking:1976ra,Fiola:1994ir}.  

As a continuation of the work \cite{Markkanen:2016jhg} here we explore the gravitational implications from a particular coarse grained density matrix: the cosmological event horizon of de Sitter space splits the Universe into observable and unobservable patches essentially identically to a black hole, which motivates us to disregard all information contained beyond the horizon. By using this density matrix to calculate the expectation values via the Friedmann equations we then perform a complete 4-dimensional back reaction analysis of de Sitter space with a conformal scalar field initialized in the Bunch-Davies vacuum.

Since the event horizon of de Sitter space is an observer-dependent concept, particle creation associated with the cosmological horizon was in \cite{Gibbons:1977mu} argued to lead to an observer dependence of the back reaction and hence of the metric of spacetime. There it was further concluded that the energy-momentum tensor sourcing the semi-classical Einstein equations cannot be defined in an observer-independent manner. Although during the time of writing, in \cite{Gibbons:1977mu} the derivation of this energy-momentum tensor was 'in preparation' the calculation to the best of our knowledge does not exist in literature. To a degree this gap is filled by the current work and since the implemented coarse graining prescription is defined by the cosmological horizon of de Sitter space it possesses the observer dependence put forward in \cite{Gibbons:1977mu}.

We will also discuss our results in the framework of horizon thermodynamics and provide a complete physical picture of particle creation in de Sitter space leading to a consistent definition of the differential of internal energy. With this picture we are able to formulate the first law of thermodynamics in de Sitter space, which is known to be problematic \cite{Padmanabhan:2002ji}, with which we show how the first principle result admits an alternate derivation by using only thermodynamic concepts. 

We emphasize that as far as horizons and particle creation are concerned there is no compelling reason to assume the arguments given not to apply also for non-conformal scalar fields, fermions and vector fields.

A 2-dimensional calculation of this mechanism was initially presented in \cite{Markkanen:2016vrp} but here we will provide much more detail, perform also the 4-dimensional calculation and make the connection to horizon thermodynamics.

Our conventions are (+,+,+) \cite{Misner:1974qy} and $c\equiv k_{B}\equiv\hbar\equiv1$.

\section{The set-up}
In $n$-dimensions the matter action for a conformally coupled scalar field is written as
\ee{S_m = -\int d^nx\sqrt{-g}~\bigg[\f{1}{2}\nabla_\mu \phi\nabla^\mu \phi+\f{\xi_n}{2} R\phi^2\bigg]\,, \label{eq:act1}} with \ee{\nonumber\xi_n=\f{(n-2)}{4(n-1)}\,,}
where $g$ is the determinant of the metric and $R$ the scalar curvature. The equation of motion for the scalar field is
\ee{\big(\Box-\xi_n R\big){\phi}=0\,, \label{eq:eom}}
where ${\sqrt{-g}}\,\Box=\partial_\mu(\sqrt{-g}\,\partial^\mu)$. The gravitational action is given by the usual Einstein-Hilbert Lagrangian supplemented by the cosmological constant term $\Lambda$
\ee{S_g= \int d^nx\sqrt{-g}~\f{1}{16\pi G}\bigg[R-2\Lambda\bigg]\,,\label{eq:EH}}
which along with (\ref{eq:act1}) leads to Einstein's equation
\ee{G_{\mu\nu}+g_{\mu\nu}\Lambda=\f{1}{M_{\rm pl}^2}\,T_{\mu\nu}\,,\label{eq:eee}} where $T_{\mu\nu}$ is the energy-momentum tensor of the scalar field
\ea{ {T}_{\mu\nu}&=-\f{g_{\mu\nu}}{2}\nabla_\rho {\phi}\nabla^\rho {\phi}+\nabla_\mu {\phi}\nabla_\nu {\phi} \nonumber \\&+\xi_n\big[G_{\mu\nu}-\nabla_\mu\nabla_\nu+g_{\mu\nu}\Box\big] {\phi}^2\label{eq:munu}\,,}
and we have defined the reduced Planck mass $M_{\rm pl}\equiv (8\pi G)^{-1/2}$. 

In four dimensions a spacetime that is expanding in a homogeneous and isotropic manner and has flat spatial sections can be expressed in the form of an FLRW line element with
\ea{ds^2&=-dt^2+a^2(t)d\mathbf{x}^2\nonumber \\ &\equiv-(dx^0)^2+a^2\big[(d{x}^1)^2+(d{x}^2)^2+(d{x}^3)^2\big]
\label{eq:FLRW}\,,} which as far as we know describes the observable Universe to good accuracy. For the line element (\ref{eq:FLRW}) the Einstein equation (\ref{eq:eee}) reduces to the  Friedmann equations\,
\ea{
\begin{cases}\phantom{-(}3H^2M_{\rm pl}^2&= \rho_m+\rho_\Lambda\\ -(3H^2+2\dot{H})M_{\rm pl}^2 &= p_m+p_\Lambda \end{cases}\,,\label{eq:e}}
where we made use of the standard definitions for the Hubble rate and the energy and pressure densities for the scalar field and the vacuum energy contributions as
\ee{H\equiv \f{\dot{a}}{a}\,;\quad \rho_m \equiv T_{00}\,;\quad p_m \equiv T_{ii}/a^2\,;\quad \rho_\Lambda\equiv\Lambda M_{\rm pl}^2\,,}
with $\rho_\Lambda\equiv -p_\Lambda$. Summing together the Friedmann equations we obtain a self-consistent evolution equation for $H$ and the scale factor $a$
\ee{-2\dot{H}M^2_{\rm pl}=\rho_m+p_m\,.\label{eq:Hev}}
The above is a very important relation in FLRW spaces as it allows one to study the back reaction of an arbitrary matter distribution onto the Hubble rate.

The Friedmann equations (\ref{eq:e}) can also be generalized to include quantum effects and hence be used in the difficult task of determining the back reaction in a quantized theory. This can be done by using the expectation values of the renormalized quantum energy-momentum
\ee{ T_{\mu\nu}\equiv \langle \hat{T}_{\mu\nu}\rangle-\delta T_{\mu\nu}\,,} where $\delta T_{\mu\nu}$ contains the counter terms, as the source term. This approach is of course not fully quantum but rather semi-classical since the spacetime metric is not quantized. However, in cases where the curvature of spacetime is not extreme this approach is expected to give reliable results \cite{Birrell:1982ix,Parker:2009uva} and it is the framework to be adopted in this work.

\section{General features of back reaction in de Sitter space}
\label{sec:gen}
As we  elaborate in section \ref{sec:coors}, when de Sitter space is parametrized in terms of an expanding FLRW metric it can be described with an exponential scale factor $a=e^{Ht}$ with $\dot{H}=0$. From this follows an important consistency condition: a strictly constant Hubble rate under back reaction in de Sitter space (\ref{eq:Hev}) implies that any classical or quantum matter distribution must satisfy \ee{\rho_m + p_m=0\,,} i.e. it must have the same equation of state as $\rho_\Lambda$. For a conformal theory this fact alone can be shown to lead to an incompatibility of having $\dot{H}=0$ and any non-zero energy density for the matter component.

For completeness we consider first the case of $n$-dimensions. A conformally coupled classical field with the action (\ref{eq:act1}) has a vanishing trace
\ee{{T_\mu}^\mu=0
\qquad \Leftrightarrow\qquad \rho_m+p_m=\f{n}{n-1}\rho_m\,,}
which can be shown from (\ref{eq:munu}) with the help of the equation of motion (\ref{eq:eom}). So at least classically, a conformal theory satisfies $\rho_m+p_m=0$ only when $\rho_m=0$.

In the quantized case the previous argument is made more complicated by the counter term contribution $\delta T_{\mu\nu}$, which leads to an anomalous trace \cite{Duff:1977ay,Deser:1976yx,Capper:1975ig,Capper:1974ic} and in de Sitter space in even dimensions gives ${T_\mu}^\mu= -\delta{T_\mu}^\mu\neq 0$. This however does not introduce a significant modification compared to the classical case discussed above. 

Any consistent prescription of renormalization of a quantum field theory one should in principle be able to express as a redefinition of the constants of the original Lagrangian. In curved space this means that generically local curvature terms such as a term $\propto R^2$ are required in the Lagrangian by consistency  \cite{Birrell:1982ix}. As explained in detail in \cite{Markkanen:2016aes}, the counter terms inherit the high degree of symmetry of de Sitter background such that all allowed counter terms for the energy-momentum tensor satisfy $\delta T_{00}=-\delta T_{ii}/a^2$, which essentially means that the counter terms in de Sitter space may be obtained by a redefinition of the cosmological constant. From this it follows that the counter terms and hence the conformal anomaly play no role in the dynamical equation (\ref{eq:Hev}) since  
\ea{\rho_m+p_m&=-\underbrace{(\delta T_{00}+\delta T_{ii}/a^2)}_{=0}+\langle \hat{T}_{00}\rangle+\langle \hat{T}_{ii}\rangle/a^2\nonumber \\&=\f{n}{n-1}\langle \hat{T}_{00}\rangle\,.\label{eq:rpq}}
The above equation also implies that for this argument to hold for a conformal theory there should be no divergences in the energy density. When using a covariant regularisation scheme such as dimensional regularization one may easily understand this to be true since in a conformal theory all scales should drop out from the vacuum terms and divergences with the correct dimensions cannot be generated. This one can easily verify with the results of \cite{Markkanen:2016jhg}. Generically, depending on the choice of regularization some divergences may have to be subtracted by hand \cite{Maggiore:2010wr2}. In what follows we will define the quantity $\rho^S_m$ to be the finite state dependent contribution to the quantum energy density i.e. the contribution that cannot be absorbed to a redefinition of the cosmological constant. Essentially, $\rho^S_m$ contains the non-trivial physical contribution of a given state and it is the only quantity needed for determining the back reaction for a conformal theory in de Sitter\footnote{When dimensional regularization is used one simply has $\langle \hat{T}_{00}\rangle=\rho^S_m$.}.

Focussing on the 4-dimensional case, the semi-classical back reaction from (\ref{eq:Hev}) now reads\footnote{We note that \ref{eq:import} coincides with equation (3.23) of \cite{Markkanen:2016jhg}, where a more detailed derivation may be found.}
\ee{-2M^2_{\rm pl}\dot{H}=\rho_m+p_m=\f{4}{3}\rho^S_m\,.\label{eq:import}} This allows us to write a set of four conditions that cannot be simultaneously satisfied:
\begin{itemize}
\item[(1)] A conformally coupled theory
\item[(2)] A FLRW line element
\item[(3)] $\rho^S_m\neq0$
\item[(4)] $\dot{H}=0$
\end{itemize}

For example, in a thermal state where $\rho^S_m$ is non-zero with a black-body spectrum the above conditions immediately imply that the Hubble rate $H$ cannot be strictly constant. More generally, if there exists a density of conformal matter that may be thought to contain any entropy it must be in a non-vacuous state with $\rho^S_m\neq0$ since the vacuum configuration is described by pure state which has strictly zero entropy, see \cite{Markkanen:2016jhg} for more discussion.

\section{De Sitter space in FLRW and static coordinates}
\label{sec:coors}
The topic of de Sitter space in various coordinates has been extensively studied in literature, for example see chapter 5 of \cite{Birrell:1982ix} for a detailed discussion.

The complete $n$-dimensional de Sitter manifold can be understood as all points contained in the $n$-dimensional hyperboloid embedded in $(n+1)$-dimensional Minkowski space. The 4-dimensional de Sitter space can then be expressed in terms of a 5-dimensional Minkowski line element
\ee{ds^2=-(dy^0)^2+(dy^1)^2+(dy^2)^2+(dy^3)^2+(dy^4)^2\,,}
with the constraint expressed in terms of some constant $H$
\begin{figure}
\begin{center}
\hspace{1.5cm}
\includegraphics[width=0.33\textwidth,origin=c,trim={0cm 19cm 10cm 0cm},clip]{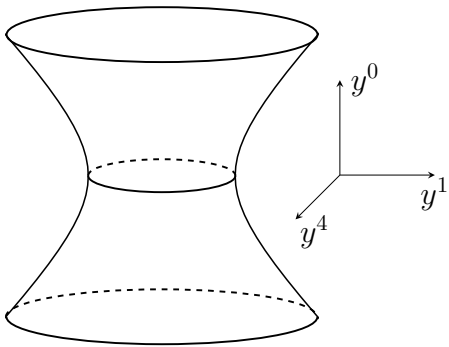}
\end{center}
\caption{The complete de Sitter manifold in four dimensions, where the $y^2$ and $y^3$ coordinates have been suppressed.\label{fig:dS}}
\end{figure}
\ee{-(y^0)^2+(y^1)^2+(y^2)^2+(y^3)^2+(y^4)^2=H^{-2}\,,}
which is depicted in Fig. \ref{fig:dS}.
The flat FLRW form of the de Sitter line element can be obtained by using the parametrization
\ea{\begin{cases}y^0&=H^{-1}\sinh(Ht)+(H/2)\vert\mathbf{x}\vert^2e^{Ht}\\  y^i&=e^{Ht}x^i \\
y^4&=H^{-1}\cosh(Ht)-(H/2)\vert\mathbf{x}\vert^2e^{Ht}\end{cases}\,,\label{eq:FLRW0}}
giving
\ee{ds^2=-dt^2+e^{2Ht}d\mathbf{x}^2\,,\label{eq:FLRW3}}
where $t \in [-\infty,\infty]$ and $x^i\in [-\infty,\infty]$. The spherical polar coordinates are defined in the usual manner in terms of the radial, polar and azimuthal coordinates $r$, $\theta$ and $\varphi$, respectively
\ea{x^i&=r\hat{n}^i\,;\qquad \hat{n}^i=(\sin\theta \cos\varphi,\sin\theta \sin\varphi,\cos\theta)\,;\nonumber \\r&\in [0,\infty]\,,\quad \theta\in [0,\pi]\,,\quad \varphi\in [0,2\pi]\,.}
\begin{figure}
\begin{center}
\includegraphics[width=0.22\textwidth,angle=90,origin=c,trim={0.2cm 0cm 1.88cm 0cm},clip]{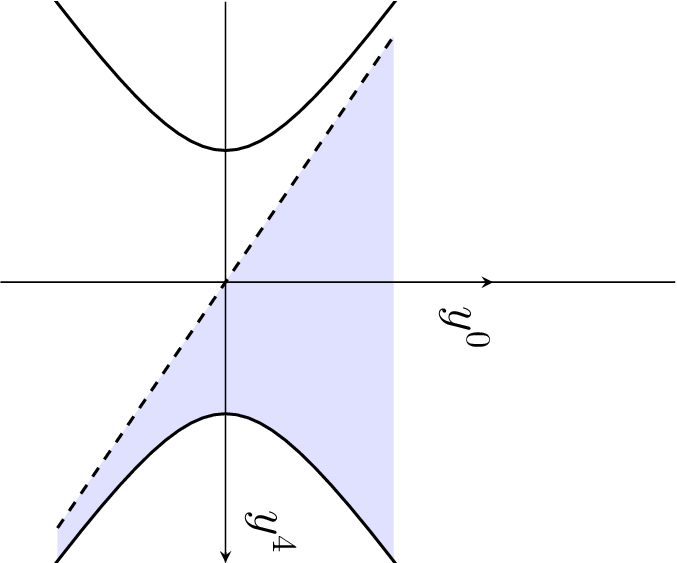}
\hspace{2cm}
\includegraphics[width=0.22\textwidth,angle=90,origin=c,trim={0.2cm 0cm 1.88cm 0cm},clip]{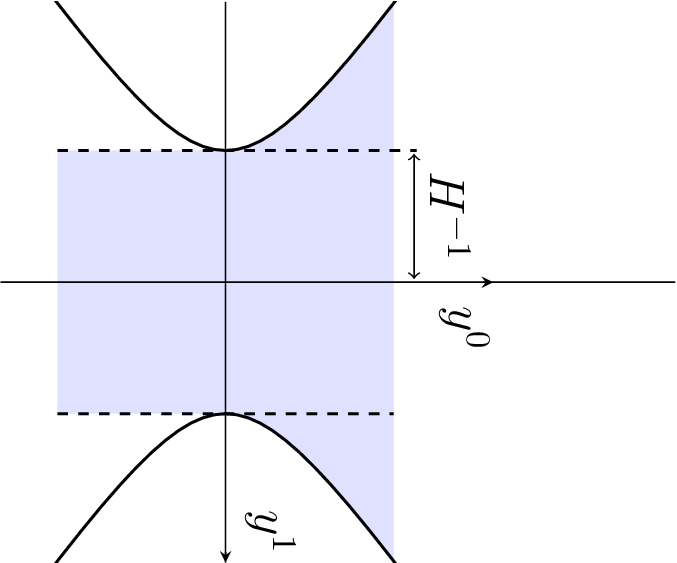}
\end{center}
\caption{Projection of the 2-dimensional de Sitter space, $-(y^0)^2+(y^1)^2+(y^4)^2=H^{-2}$, as covered by the expanding FLRW coordinates (\ref{eq:FLRW0}) denoted with blue on the $(y^4,y^0)$-plane (top) and the $(y^1,y^0)$-plane (bottom).  \label{fig:FLRW}} 
\end{figure}From the relations (\ref{eq:FLRW0}) we see that in these coordinates $y^0+y^4 \geq 0$, which is not a property of the complete de Sitter manifold. This means that (\ref{eq:FLRW0}) do not cover the entire manifold, but only half of it. This is illustrated for the 2-dimensional case in Fig. \ref{fig:FLRW}. The coordinates covering the other half with $y^0+y^4 \leq 0$ can be obtained by setting $Ht\rightarrow-Ht +i\pi$ in (\ref{eq:FLRW0}), which also  leads to a FLRW line element, but with an exponentially contracting scale factor.

An important feature of the coordinates (\ref{eq:FLRW0}) is that they cover regions of spacetime a local observer would not be able to interact with. This can be shown by calculating the maximum physical distance at a time $t$ that can be reached by a ray of light emanating from the origin
\ee{e^{Ht}r_\infty=e^{Ht}\int_{t}^\infty \f{dt'}{e^{Ht'}}=H^{-1}\,,\label{eq:hor}} 
which of course is cut-off by the cosmological event horizon, which from now on we will simply call the horizon.
\begin{figure}
\begin{center}
\includegraphics[width=0.22\textwidth,angle=90,origin=c,trim={0.2cm 0cm 1.88cm 0cm},clip]{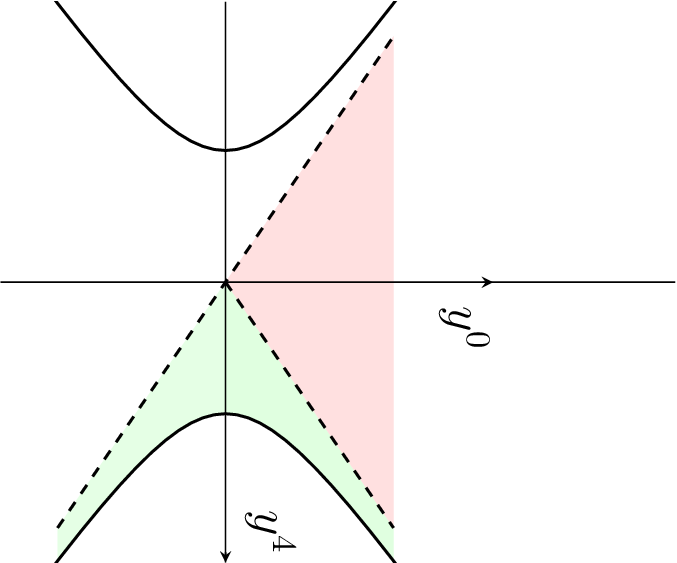}
\hspace{2cm}
\includegraphics[width=0.22\textwidth,angle=90,origin=c,trim={0.2cm 0cm 1.88cm 0cm},clip]{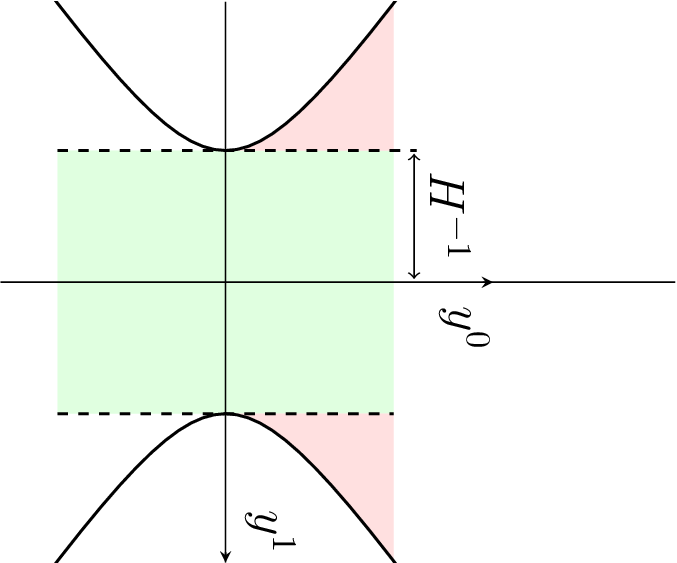}
\end{center}
\caption{Projection of the 2-dimensional de Sitter space, $-(y^0)^2+(y^1)^2+(y^4)^2=H^{-2}$, as covered by the static-type coordinates (\ref{eq:staA}) and (\ref{eq:staB}) denoted with green and red, respectively, on the $(y^4,y^0)$-plane (top) and the $(y^1,y^0)$-plane (bottom).  \label{fig:STA}} 
\end{figure}

In the static parametrization of de Sitter space only the spacetime inside the horizon is covered
\ea{\begin{cases}y^0&=(H^{-2}-\ubar{r}_\text{\tiny\it A}^2)^{1/2}\sinh(H \ubar{t}_\text{\tiny \it A})\\ y^i&=\ubar{r}_\text{\tiny\it A}\hat{\ubar{n}}^i\\
y^4&=(H^{-2}-\ubar{r}_\text{\tiny \it A}^2)^{1/2}\cosh(H \ubar{t}_\text{\tiny \it A})\end{cases}\,,\label{eq:staA}}
 which is explicitly borne out by a singularity in the line element
\ea{ds^2&=-\big[1-(H\ubar{r}_\text{\tiny \it A})^2\big]d\ubar{t}_\text{\tiny \it A}^2+\big[1-(H\ubar{r}_\text{\tiny \it A})^2\big]^{-1}d\ubar{r}_\text{\tiny \it A}^2 \nonumber \\&+\ubar{r}_\text{\tiny \it A}^2\big(d\ubar{\theta}_\text{\tiny \it A}^2+\sin^2 \ubar{\theta}_\text{\tiny \it A}d\ubar{\varphi}^2\big)\,,\label{eq:sta}}
where $\ubar{t}_\text{\tiny \it A} \in [-\infty,\infty]$ and $\ubar{r}_\text{\tiny \it A}\in [0,1/H]$, and throughout we will use underlines to distinguish the static coordinates from the FLRW ones. In particular, $\ubar{x}^i$ describes physical distance and ${x}^i$  comoving distance. We can easily verify in the static coordinates that the time it takes for a light ray to reach the horizon diverges, in agreement with (\ref{eq:hor}). The patch covered by (\ref{eq:staA}) for the 2-dimensional case is shown in Fig. \ref{fig:STA} as the green region.

The reason for the $A$ subscript in (\ref{eq:staA}) is that we also need coordinates covering the spacetime outside the horizon. These we parametrize with
\ea{\begin{cases}y^0&=(\ubar{r}_\text{\tiny \it B}^2-H^{-2})^{1/2}\cosh(H \ubar{t}_\text{\tiny \it B})\\ y^i&=\ubar{r}_\text{\tiny \it B}\hat{\ubar{n}}^i\\
y^4&=(\ubar{r}_\text{\tiny \it B}^2-H^{-2})^{1/2}\sinh(H \ubar{t}_\text{\tiny \it B})\end{cases}\,,\label{eq:staB}}
where $\ubar{t}_\text{\tiny \it B} \in [-\infty,\infty]$ and $\ubar{r}_\text{\tiny \it B}\in [1/H,\infty]$ and the line element is precisely as in (\ref{eq:sta}), but with $A\rightarrow B$. Note that in the 2-dimensional case studied in section \ref{sec:2dim} one also has a second patch beyond the horizon, which we denote with $C$. The region covered by (\ref{eq:staB}) is shown as the red region in Fig. \ref{fig:STA}.
It is worth pointing out that the coordinates (\ref{eq:staA}) and (\ref{eq:staB}) significantly resemble the Kruskal--Szekeres parametrization of the Schwarzschild black hole. Using the combination of the coordinates (\ref{eq:staA}) and (\ref{eq:staB}) one may cover the same patch as the FLRW system (\ref{eq:FLRW0}) as can be seen by comparing Fig. \ref{fig:FLRW} and Fig. \ref{fig:STA}. The regions covered separately by coordinates of type (\ref{eq:staA}) and (\ref{eq:staB}) we will refer to as regions $A$ and $B$ or the static patches although this is not strictly speaking a good characterization for (\ref{eq:staB}): since $[1-(H\ubar{r}_\text{\tiny \it B})^2\big]<0$, $\ubar{r}_\text{\tiny \it B}$ is in fact the time coordinate and hence the metric (\ref{eq:sta}) is explicitly time dependent beyond the horizon. 
\section{The coarse grained energy-momentum tensor}
\label{eq:sec3}
In this section we assume a strictly de Sitter background throughout with\ee{a=e^{Ht}\,.}

Up until now we have written the expectation value of the energy-momentum tensor symbolically as $\langle \hat{T}_{\mu\nu}\rangle$, without specifying how it is to be derived. As mentioned in the introduction, our method for calculating the expectation value will deviate from what is generally used for the semi-classical prescription \cite{Birrell:1982ix}. Before this important topic we need to however cover some basic features.

First, we will adopt the cosmologically motivated choice where de Sitter space is described in terms of the expanding FLRW coordinates used for example when calculating the cosmological perturbations from inflation.
This is a consistent approach, since for an observer at rest with the expanding FLRW coordinates the contracting patch is not accessible.

Next we need to define the specific state to be used as the initial condition. In the black hole context making the physical choice for the vacuum is an essential ingredient for the understanding of the evaporation process, see \cite{Unruh:1976db} for important pioneering work and \cite{Padmanabhan:2003gd} for a clear discussion. Again conforming to the usual choice made in inflationary cosmology, we will use the Bunch-Davies vacuum \cite{Chernikov:1968zm,BD} as the quantum sate. A compelling motivation for this choice comes from the fact that the Bunch-Davies vacuum is an attractor state in de Sitter space \cite{Markkanen:2016aes}, provided we make the natural assumption that the leading divergences of the theory coincide with those in flat space \cite{Allen:1985ux,Bros:1995js}. 

A very important feature of the Bunch-Davies vacuum is that it covers the entire FLRW de Sitter patch and hence extends also to regions that would be hidden behind the horizon. This is a natural requirement for an initial condition in a case where the Universe was not always dominated by vacuum energy, which from the cosmological point of view is well-motivated: for example, the Universe may start out as radiation or matter dominated and only at late times asymptotically approach the exponentially expanding de Sitter space as in the $\Lambda$CDM model. This will turn out to be crucial for our calculation.

For deriving the Bunch-Davies vacuum in the FLRW coordinates it proves convenient to make use of the conformal time coordinate
\ee{\,d\eta = \f{dt}{a}\quad\Rightarrow\quad \eta = \f{-1}{aH}\quad\Rightarrow\quad ds^2=a^2\big[-d\eta^2+d\mathbf{x}^2\big]\,,\label{eq:conf}} with which the equation of motion (\ref{eq:eom}) becomes
\ee{\bigg[\partial_\eta^2+(n-2)\f{a'}{a}\partial_\eta-a^2\partial_i\partial^i+\f{n-2}{4(n-1)}a^2R\bigg]\hat{\phi}=0\,,\label{eq:eommi}}
where $a'\equiv \partial a / \partial\eta$ and $a^2R=2(n-1)a''/a+(n-1)(n-4)(a'/a)^2$.
The solutions can be written as the mode expansion
\ee{\hat{\phi}=\int d^{n-1}\mathbf{k}\left[\hat{a}_\mathbf{k}^{\phantom{\dagger}}u^{\phantom{\dagger}}_{\mathbf{k}}+\hat{a}_{\mathbf{k}}^\dagger u^{*\phantom{\dagger}}_{{\mathbf{k}}}\right]\,\label{eq:adsol2}\, ,}
with $k\equiv \vert \mathbf{k}\vert$, with the commutation relations  $[\hat{a}_{\mathbf{k}}^{\phantom{\dagger}},\hat{a}_{\mathbf{k}'}^\dagger]=\delta^{(n-1)}(\mathbf{k}-\mathbf{k}'),~~[\hat{a}_{\mathbf{k}}^{\phantom{\dagger}},\hat{a}_{\mathbf{k}'}^{\phantom{\dagger}}]=[\hat{a}_{\mathbf{k}}^{{\dagger}},\hat{a}_{\mathbf{k}'}^\dagger]=0$ and the modes
\ee{u^{\phantom{\dagger}}_{\mathbf{k}}=\f{1}{\sqrt{(2\pi)^{n-1}a^{n-2}}}\f{1}{\sqrt{2k}}e^{-i (k\eta-\mathbf{k\cdot\mathbf{x}})}\, ,\label{eq:bog}}
which define the Bunch-Davies vacuum state $\vert 0 \rangle$ via
\ee{\hat{a}_{\mathbf{k}}\vert 0 \rangle=0\, .\label{eq:vac}}

The Klein-Gordon inner product between two solutions to the equation of motion $\phi_1$ and $\phi_2$ can be defined in terms of a spacelike hypersurface $\Sigma$, a future oriented unit normal vector $n^\mu$ and the induced spatial metric $\gamma_{ij}$ as
\ee{\big(\phi_1,\phi_2\big)=-i\int_\Sigma d^{n-1}\mathbf{x}\sqrt{\gamma}n^\mu\,\phi_1^{\phantom{*}}\overset{\leftrightarrow}{\nabla}_\mu\phi^*_2\,;\quad\overset{\leftrightarrow}{\nabla}_\mu=\overset{\rightarrow}{\nabla}_\mu-\overset{\leftarrow}{\nabla}_\mu\,.\label{eq:norma}}
It is easy to show in the conformal coordinates (\ref{eq:conf}) that using the vector $\partial_\eta$ for normalization gives $n^\eta=a^{-1}, n^i=0$ and $\sqrt{\gamma} = a^{n-1}$ with which the inner product takes the form
\ee{\big(\phi_1,\phi_2\big)=-i\int d^{n-1}\mathbf{x}\,a^{n-2}\,\phi_1^{\phantom{*}}\overset{\leftrightarrow}{\nabla}_\eta\phi^*_2\,,}
and that the expansion (\ref{eq:adsol2}) in terms of the Bunch-Davies modes (\ref{eq:bog}) is properly normalized,
\ee{\big(u^{\phantom{\dagger}}_{\mathbf{k}},u^{\phantom{\dagger}}_{\mathbf{k}'}\big)=\delta^{(n-1)}(\mathbf{k}-\mathbf{k}')\,.}

For completeness we also show the derivation for the Bunch-Davies modes in the spherical coordinates in four dimensions. Using an ansatz
\ee{\psi_{\ell m k}=f_{\ell k}(r)Y^m_\ell \f{e^{-ik\eta}}{a}\label{eq:ans4}\,,}
where the $Y_\ell^m$ are the spherical harmonics normalized according to 
\ea{\int d\Omega \,Y_\ell^m \, Y_{\ell'}^{m'}{}^*&=\int_{\theta=0}^\pi\int_{\varphi=0}^{2\pi}d\theta d\varphi \sin \theta \,Y_\ell^m \, Y_{\ell'}^{m'}{}^* \nonumber \\& =\delta_{\ell\ell'}\, \delta_{mm'}\,,}
the equation of motion (\ref{eq:eommi}) reduces to a purely radial equation
\ee{\bigg\{r^{-2}\f{\partial}{\partial r}\Big(r^2\f{\partial}{\partial{{r}}}\Big)+{k^2} -\f{\ell(\ell+1)}{r^2}\bigg\}f_{\ell k}({r})=0\,,}
which has solutions expressible as linear combinations of the spherical Bessel functions $j_\nu(x)$. Writing the inner product in spherical coordinates 
\ee{\big(\phi_1,\phi_2\big)=-i\int d\Omega\int_0^\infty dr\,r^2\,a^{2}\,\phi_1^{\phantom{*}}\overset{\leftrightarrow}{\nabla}_\eta\phi^*_2\,,} 
and making use of the orthogonality properties of the Bessel functions allows one to derive the properly normalized positive frequency modes
\ee{\psi_{\ell m k}=\sqrt{\f{k}{\pi}}j_\ell(kr)Y^m_\ell \f{e^{-ik\eta}}{a}\label{eq:ans3}\,.}
The spherical modes (\ref{eq:ans3}) provide another representation for the scalar field and the Bunch-Davies vacuum via 
\ee{\hat{\phi}=\sum_{\ell =0}^\infty\sum_{m=-\ell}^\ell\int_0^\infty dk\Big[\psi^{\phantom{\dagger}}_{\ell m k}\hat{a}^{\phantom{\dagger}}_{\ell m k}+\psi^{*\phantom{\dagger}}_{\ell m k}\hat{a}^\dagger_{\ell m k}\Big]\label{eq:sphephi}}
and
\ee{\hat{a}^{\phantom{\dagger}}_{\ell m k}|0\rangle=0\,.\label{eq:vac2}}
The equivalence of the states as defined by (\ref{eq:vac}) and (\ref{eq:vac2}) can be demonstrated for example by showing that the Wightman function as defined by the two states coincides, which can be easily done with the help of the Rayleigh or plane wave expansion.

In order to obtain well-defined quantum expectation values we must define a regularization and a renormalization prescription for the ultraviolet divergences. Perhaps the most elegant way would be to analytically continue the dimensions to $n$ and redefine the constants of the original action to obtain physical results. Dimensional regularization does have a drawback however, which is that consistency requires one to calculate everything in $n$ dimensions, which is surely more difficult than to perform the calculation in 2 dimensions, for example. For our purposes the most convenient choice is the adiabatic subtraction technique \cite{Parker:1974qw1, Bunch:1980vc,Parker:1974qw}, which in the non-interacting case is a consistent and a covariant approach \cite{Markkanen:2013nwa}, but where no explicit regularization is needed as the counter terms can formally be combined in the same integral as the expectation value. This also allows us to use a strictly 2- or a 4-dimensional theory.

\subsection{Tracing over the unobservable states}
\label{sec:tr}
So far our approach has followed standard lines. If in some given state $\vert\Psi\rangle$  we were to calculate the relevant expectation values $\langle\hat{T}_{\mu\nu}\rangle \equiv \langle \Psi\vert\hat{T}_{\mu\nu}\vert\Psi\rangle$ and use them as the sources in the Friedmann equations we would obtain a result that exponentially fast approaches a configuration with $\dot{H}=0$ and conclude that de Sitter space is a stable solution also when back reaction is taken into account. This is a manifestation of the de Sitter invariance and the attractor nature of the Bunch-Davies vacuum and true as well for the non-conformal case \cite{Markkanen:2016aes}. However, as discussed in section \ref{sec:coors} the horizon in de Sitter space splits the FLRW manifold into two patches only one of which is visible to a local observer. This is very much analogous to how a black hole horizon blocks the observational access of an observer outside the horizon \cite{Hawking:1976ra,Gibbons:1977mu}. Here is where our approach will differ from what is traditionally done in semi-classical gravity: following \cite{Markkanen:2016jhg} when calculating $\langle\hat{T}_{\mu\nu}\rangle$ we will use a prescription where we average over only those states that are 
inside the horizon and thus observable. We note that quite generally coarse graining a state is expected to bring about a qualitative change in the results since it often leads to a violation of de Sitter invariance \cite{Markkanen:2016jhg}. 


A configuration where a state is not completely observationally accessible can be described in terms of an open quantum system. For more discussion, see for example the textbook \cite{Calzetta:2008iqa}. If we assume that the quantum state $\vert \Psi\rangle$  can be written as a product of orthonormal states $\vert n,A\rangle$ of the observable system and $\vert n,B\rangle$ of the unobservable environment as
\ee{\vert\Psi\rangle= \sum_n p_n\vert n,A\rangle\vert n,B\rangle\,,}
we can express expectation values with a coarse grained density matrix $\hat{\rho}$
\ee{\langle\hat{O}\rangle\equiv{\rm Tr}\big\{\hat{O}\hat{\rho}\big\}\,,}
where the density matrix $\hat{\rho}$ is obtained by neglecting or tracing over the unobservable states
\ea{\hat{\rho}&\equiv {\rm Tr}_\text{\it B}\big\{\vert \Psi \rangle\langle\Psi\vert\big\} \nonumber \\&=\sum_m\langle B,m\vert \bigg\{\Big[\sum_n p_n\vert n,A\rangle\vert n,B\rangle\Big]\nonumber \\&\times\Big[\sum_{n'} p_{n'}^* \langle B,n'\vert\langle A,n'\vert\Big]\bigg\}\vert m,B\rangle\nonumber\,,} leading to\ee{\Leftrightarrow\quad \hat{\rho}=\sum_m\vert p_m\vert^2\vert m,A\rangle\langle A,m\vert\,.\label{eq:cg}}
If in the state $\vert \Psi\rangle$ there is entanglement between the observable states and the unobservable states we coarse grain over, the initially pure quantum state becomes mixed and the Von Neumann entropy of the density matrix will be non-zero
\ee{\hat{\rho}^2\neq\hat{\rho}\quad\Leftrightarrow\quad -{\rm Tr}\big(\hat{\rho}\log\hat{\rho}\big)>0\,,\label{eq:en}}
signalling that part of the information of the initial state $\vert\Psi\rangle$ is lost or unobservable. When coarse graining leads to entropy increase/information loss it is a generic feature that the expectation values will not remain the same \cite{Markkanen:2016jhg}. For example for the energy-momentum tensor one would expect to have
\ee{
{\rm Tr}\big\{\hat{T}_{\mu\nu}\hat{\rho}\big\}\quad{\neq}\quad \langle  \Psi |\hat{T}_{\mu\nu}| \Psi\rangle\,,}
implying that the coarse grained system has a different gravitational response compared to the un-coarse grained case. 

Importantly, the Bunch-Davies vacuum in de Sitter space before coarse graining is a zero entropy state, but as explained covers also regions that are hidden from a local observer. If tracing over the unobservable states leads to a non-zero entropy it also suggests the presence of a non-zero energy density, which in light of the arguments given in section \ref{sec:gen} implies $\dot{H}\neq0$ and gives an important link between loss of information from coarse graining and a potentially non-trivial back reaction in our prescription.

Our choice of neglecting the unobservable states from the expectation values can be motivated as follows. First of all it is a standard procedure in branches of physics where having only partial observable access to a quantum state is a typical feature. An important example is the decoherence program: without an unobservable environment the quantum-to-classical transition does not take place \cite{Zurek:2003zz}. Neglecting unobservable information is crucial also for the inflationary paradigm: in order to obtain the correct evolution of large scale structure as seeded by the inherently quantum fluctuations from inflation one must calculate the gravitational dynamics from the classicalized i.e. coarse grained energy-momentum tensor \cite{Polarski:1995jg}. Perhaps most importantly, the energy-momentum tensor one obtains after neglecting the unobservable states corresponds to what an observer would actually measure and in this sense has clear physical significance. 

A profound feature of our prescription is that since the horizon in de Sitter space is an observer dependent quantity, so is then the back reaction itself. Although a rather radical proposition, this does not imply an immediate inconsistency. After all, observer dependence is a ubiquitous feature in general -- and even special -- relativity. Furthermore, the well-known observer dependence of the concept of a particle in quantum theories on curved backgrounds was argued to lead to such a conclusion already in the seminal work \cite{Gibbons:1977mu}.

Although our prescription of using a coarse grained energy-momentum tensor as the source term for semi-classical gravity deviates from the standard approach making use of $\langle\hat{T}_{\mu\nu}\rangle \equiv \langle \Psi\vert\hat{T}_{\mu\nu}\vert\Psi\rangle$, we would like to emphasize that at the moment there is no method for conclusively determining precisely which object is the correct one \cite{Birrell:1982ix}. This stems from the fact that semi-classical gravity is not a complete first principle approach, but rather an approximation for describing some of the gravitational implications from the quantum nature of matter. Before a full description of quantum gravity is obtained it is likely that this state of affairs will remain.

Tracing over inaccessible environmental states that are separated by a sharp boundary from the accessible ones generically leads to divergent behaviour close to the boundary. This is encountered for example in the context of black hole entropy \cite{tHooft:1984kcu} and entanglement entropy in general \cite{Srednicki:1993im,Bombelli:1986rw}. Although by introducing a cut-off or a smoothing prescription well-defined results can be obtained \cite{Fiola:1994ir}, there is valid suspicion of the applicability of the semi-classical approach when close to the horizon. However, we can expect reliable results at the limit when the horizon is far away. At this limit there exists a natural expansion in terms of physical distance in units of the horizon radius, or more specifically in terms of the dimensionless quantities $H\ubar{x}^i$, in the notation of section \ref{sec:coors}. The neglected terms we will throughout denote as $\mathcal{O}(H\ubar{x})$. This limit can be expressed equivalently as being far away from the horizon or close to the center of the Hubble sphere and can equally well be satisfied when $H$ is large such as during primordial inflation or when it is very small as it is during the late time Dark Energy dominated phase we are currently entering. The limit where the observer is far from the horizon is also the limit taken in the standard black hole analysis \cite{Hawking:1974sw}.


\subsection{Two dimensions}
\label{sec:2dim}
For completeness we first go through the steps of the 2-dimensional argument presented in \cite{Markkanen:2016vrp}, before proceeding to the full 4-dimensional derivation.

The various coordinate systems in de Sitter space in two dimensions can be expressed analogously to what was discussed in four dimensions in section \ref{sec:coors}. The main modification is that since there is only one spatial coordinate there are now two horizons, at $\pm1/H$. For the patch inside the horizon in two dimensions one may use the FLRW
\ee{\begin{cases}y^0&=H^{-1}\sinh(Ht)+(H/2){x}^2e^{Ht}\\  y^1&=e^{Ht}x \\
y^4&=H^{-1}\cosh(Ht)-(H/2){x}^2e^{Ht}\end{cases}\,,\label{eq:FLRW11}}
or static coordinates \ee{\begin{cases}y^0&=(H^{-2}-\ubar{x}_\text{\tiny \it A}^2)^{1/2}\sinh(H \ubar{t}_\text{\tiny \it A})\\ y^1&=\ubar{x}_\text{\tiny \it A}\\
y^4&=(H^{-2}-\ubar{x}_\text{\tiny \it A}^2)^{1/2}\cosh(H \ubar{t}_\text{\tiny \it A})\end{cases}\,,\label{eq:FLRW1}}
giving
\ee{ds^2=-dt^2+e^{2Ht}d{x}^2\,,\label{eq:FLRW2}}
and
\ee{ds^2=-\big[1-(H\ubar{x}_\text{\tiny \it A})^2\big]d\ubar{t}_\text{\tiny \it A}^2+\big[1-(H\ubar{x}_\text{\tiny \it A})^2\big]^{-1}d\ubar{x}_\text{\tiny \it A}^2 \,,\label{eq:sta2}}
respectively, with $t,\ubar{t}_\text{\tiny \it A},x\in [-\infty,\infty]$ and $\ubar{x}_\text{\tiny \it A}\in [-1/H,1/H]$. Since in two dimensions there are two horizons there are also two patches beyond the horizon that can be covered with the FLRW coordinates or with
\ee{\begin{cases}y^0&=(\ubar{x}_\text{\tiny \it B}^2-H^{-2})^{1/2}\cosh(H \ubar{t}_\text{\tiny \it B})\\ y^1&=\ubar{x}_\text{\tiny \it B}\\
y^2&=(\ubar{x}_\text{\tiny \it B}^2-H^{-2})^{1/2}\sinh(H \ubar{t}_\text{\tiny \it B})\end{cases}\,,\label{eq:staB22}}
where $\ubar{t}_\text{\tiny \it B} \in [-\infty,\infty]$ and $\ubar{x}_\text{\tiny \it B}\in [1/H,\infty]$, and \ee{\begin{cases}y^0&=(\ubar{x}_\text{\tiny \it C}^2-H^{-2})^{1/2}\cosh(H \ubar{t}_\text{\tiny \it C})\\ y^1&=\ubar{x}_\text{\tiny \it C}\\
y^2&=(\ubar{x}_\text{\tiny \it C}^2-H^{-2})^{1/2}\sinh(H \ubar{t}_\text{\tiny \it C})\end{cases}\,,\label{eq:staB2}}
where $\ubar{t}_\text{\tiny \it C} \in [-\infty,\infty]$ and $\ubar{x}_\text{\tiny \it C}\in[-\infty,-1/H]$.

\begin{figure}
\begin{center}
\includegraphics[width=0.3\textwidth,trim={0 1.0cm 0 0},clip]{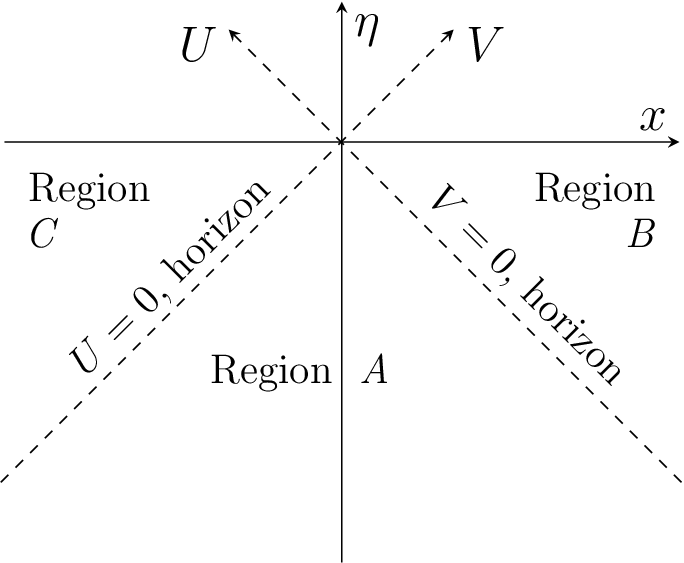}
\end{center}
\caption{The expanding FLRW patch in de Sitter space is described by $\eta \leq0$ and $-\infty<x<\infty$, and can be covered by the $A,B$ and $C$ coordinate systems defined in (\ref{eq:FLRW1}), (\ref{eq:staB22}) and (\ref{eq:staB2}).\label{fig:regs}}
\end{figure}

The relations between the various coordinate systems become quite simple when using the light-cone coordinates defined in terms of conformal time (\ref{eq:conf}) as
\ee{\begin{cases} V&=\eta+x=+e^{-Ht}\big(\ubar{x}-H^{-1}\big)\label{eq:LC0}\,,\\
U&=\eta-x=-e^{-Ht}\big(\ubar{x}+H^{-1}\big)\end{cases}\,,}
where the notation implies the same definition in all three regions $A,B$ and $C$. As is clear from the definitions (\ref{eq:LC0}) the $V$ and $U$ coordinates can also be conveniently used to split the FLRW patch in terms of the regions $A,B$ and $C$ since they vanish at the horizons $1/H$ and $-1/H$, respectively. This is illustrated in Fig. \ref{fig:regs}. 
Furthermore, in the static patches we define the tortoise coordinates
\ee{d\ubar{x}=\big[1-(H\ubar{x})^2\big]d\ubar{x}^*\quad \Rightarrow\quad \ubar{x}^*=(2H)^{-1}\log\bigg\vert\f{1+H\ubar{x}}{1-H\ubar{x}}\bigg\vert\,,\label{eq:tort}}
with $\ubar{x}^*_\text{\tiny \it A}\in [-\infty,\infty]$, $\ubar{x}^*_\text{\tiny \it B}\in [0,\infty]$ and $\ubar{x}^*_\text{\tiny \it C}\in [-\infty,0]$.  It is now a question of straightforward algebra to express of the light-cone coordinates $U$ and $V$ in terms of the static ones. The results can be summarized as
\ea{\begin{cases}V&{=}-H^{-1}e^{-Hv_{\text{\tiny \it A}}}\\
U&{=}-H^{-1}e^{-Hu_\text{\tiny \it A}}\end{cases}\,,\,{\rm Reg. }~A\,,\label{eq:coors21}}\ea{\begin{cases}V&{=}+H^{-1}e^{-Hv_{B}}\\
U&{=}-H^{-1}e^{-Hu_\text{\tiny \it B}}\end{cases}\,,\,{\rm Reg. }~B\,,\label{eq:coors22}} and\ea{\begin{cases}V&{=}-H^{-1}e^{-Hv_{C}}\\
U&{=}+H^{-1}e^{-Hu_\text{\tiny \it C}}\end{cases}\,,\,{\rm Reg. }~C\,,\label{eq:coors23}}
and it is also convenient to define light-cone coordinates with respect to the static coordinates\ee{\begin{cases}v&\,=\phantom{+}\,\ubar{t}+\ubar{x}^*\\
u&\,{=}\phantom{+}\,\,\ubar{t}-\ubar{x}^*\end{cases}\,. \label{eq:coors24}}

The core of the calculation is finding an expression for the coarse grained density matrix (\ref{eq:cg}), from which the unobservable information related to states  beyond the horizon is removed. If we assume that any possible entanglement occurs only between modes with the same momentum, the density matrix where the hidden states are traced over can be written as the product in momentum space
\ee{\hat{\rho}\equiv {\rm Tr}_{BC}\big\{\vert0 \rangle\langle0\vert\big\}=\prod_{\mathbf{k}}{\rm Tr}_{BC}\big\{\vert0_{\mathbf{k}} \rangle\langle0_{\mathbf{k}}\vert\big\}\equiv\prod_{\mathbf{k}} \hat{\rho}_{\mathbf{k}}\,,\label{eq:dmc} }
where we define $\mathbf{k}$ to be a scalar going from $-\infty$ to $\infty$ and where $\vert0_{\mathbf{k}} \rangle$ is the $\mathbf{k}$'th Fock space contribution to the Bunch-Davies vacuum, $\vert0 \rangle=\prod \!\!{\,}^{\,}_{\mathbf{k}} \vert0^{\,}_{\mathbf{k}} \rangle$.

Next we need to find an expression for the Bunch-Davies vacuum in terms of observable and unobservable states. This is obtained by relating the Bunch-Davies modes to the ones defined in the 2-dimensional static coordinates found in (\ref{eq:FLRW1}), (\ref{eq:staB22}) and (\ref{eq:staB2}). From (\ref{eq:eommi} -- \ref{eq:vac}) we see that in two dimensions the mode expansion in de Sitter space coincides with the flat space result and can be written in the light-cone coordinates (\ref{eq:LC0}) as
\ea{\hat{\phi} &\equiv\hat{\phi}_V+\hat{\phi}_U=\nonumber\\&\int_0^\infty\f{dk}{\sqrt{4\pi k}}\Big[e^{-ikV}\hat{a}^{\phantom{\dagger}}_{-k} +e^{ikV}\hat{a}^{\dagger}_{-k} +e^{-ikU}\hat{a}^{\phantom{\dagger}}_{k} +e^{ikU}\hat{a}^{\dagger}_{k}\Big]\,.\label{eq:mode}}
In (\ref{eq:mode}) we have split the quantum field to two contributions according their dependence on $U$ or $V$ since these newer mix and can be thought as separate sectors, as is evident by taking into account\footnote{Here we neglect the $k=0$ zero mode, whose quantization is a non-trivial issue in 2-dimensional field theory \cite{Coleman:1973ci}, but does not pose problems in four dimensions.} $[\hat{a}^{\phantom{\dagger}}_{-k},\hat{a}^{{\dagger}}_{k}]=0$, $V=V(v)$ and $U=U(u)$ from (\ref{eq:coors21} -- \ref{eq:coors24}). Since $\hat{\phi}_V$ is expressed only in terms of the $\hat{a}^{\phantom{\dagger}}_{-k}$ operators it consists solely of particles moving towards the left and similarly for $\hat{\phi}_U$ and the right-moving particles.

In two dimensions also the static coordinates give rise to a trivial equation of motion
\ee{\Box\hat{\phi}=0\qquad \Leftrightarrow\qquad \big(\partial_{\ubar{t}}^2-\partial_{\ubar{x}^*}^2\big)\hat{\phi}=0\,,}
but much like for the Unruh effect, we must carefully determine the correct normalization for the modes in the static patches. Namely, we need to make sure that we are consistent in terms of defining a positive frequency mode.

As discussed after equation (\ref{eq:norma}), the modes in (\ref{eq:mode}) are defined to be positive frequency in terms of the vector $\partial_\eta$ and we need to respect this definition also in the static patches $A,B$ and $C$. If we choose $\partial_{\ubar{t}^{\phantom{*}}_\text{\tiny \it A}}$, $-\partial_{\ubar{x}^*_\text{\tiny \it B}}$ and $\partial_{\ubar{x}^*_\text{\tiny \it C}}$ for $A,B$ and $C$ respectively, it is a simple matter of using the vector transformations $\partial_{\mu}=\f{\partial x^{\tilde{\alpha}}}{\partial x^\mu}\partial_{\tilde{\alpha}}$ with (\ref{eq:coors21} -- \ref{eq:coors24}) to show that
\ea{\partial_{\ubar{t}^{\phantom{*}}_\text{\tiny \it A}}&=-H\big(\eta\partial_\eta+x\partial_x\big)\label{eq:vec1}\\\label{eq:vec2}-\partial_{\ubar{x}^{{*}}_\text{\tiny \it B}}&=+H\big(x\partial_\eta+\eta\partial_x\big)\\\partial_{\ubar{x}^{{*}}_\text{\tiny \it C}}&=-H\big(x\partial_\eta+\eta\partial_x\big)\,,\label{eq:vec3}}
which with the help of Fig. \ref{fig:regs} one may see to be time-like in terms of conformal time and future-oriented in their respective regions\footnote{For example, from Fig. \ref{fig:regs} we see that in region $C$ we have $-x\geq0$ and $UV\leq0~\Leftrightarrow~ x^2\geq\eta^2$.}.

Normalizing the vectors (\ref{eq:vec1} -- \ref{eq:vec3} ) we can then from (\ref{eq:norma}) write inner products in the static patches
\ea{(\phi_1,\phi_2)_\text{\tiny \it A}&=\label{eq:ipa}-i\int d\ubar{x}^*_\text{\tiny \it A}\phi_1\overset{\leftrightarrow}{\nabla}_{\ubar{t}^{\phantom{*}}_\text{\tiny \it A}}\phi_2^*\,;\quad \,{\rm Reg. }~A\,,\\ \label{eq:ipb}(\phi_1,\phi_2)_\text{\tiny \it B}&=+i\int d\ubar{t}^{\phantom{*}}_\text{\tiny \it B}\,\phi_1\overset{\leftrightarrow}{\nabla}_{\ubar{x}^{{*}}_\text{\tiny \it B}}\phi_2^*\,;\quad \,{\rm Reg. }~B\,,\\ (\phi_1,\phi_2)_\text{\tiny \it C}&=-i\int d\ubar{t}^{\phantom{*}}_\text{\tiny \it C}\,\phi_1\overset{\leftrightarrow}{\nabla}_{\ubar{x}^{{*}}_\text{\tiny \it C}}\phi_2^*\,;\quad \,{\rm Reg. }~C\,,\label{eq:ipc}}
with which the expression for the scalar field in the static coordinates becomes
\ea{&\hat{\phi}\nonumber= \int_0^\infty\f{dk}{\sqrt{4\pi k}}\\&\times\begin{dcases}\displaystyle \Big(e^{-ik{v}_\text{\tiny \it A}}\hat{a}^{\text{\tiny \it A}}_{-k}+ {\rm H.C}\Big)+\Big(e^{-ik{u}_\text{\tiny \it A}}\hat{a}^{\text{\tiny \it A}}_{k} + {\rm H.C}\Big)\,,\, {\rm Reg. }~A\,,\label{eq:mode1}\\\displaystyle \Big(e^{+ik{v}_\text{\tiny \it \text{\tiny \it B}}}\hat{a}^{\text{\tiny \it B}}_{-k} + {\rm H.C}\Big)+\Big(e^{-ik{u}_\text{\tiny \it B}}\hat{a}^{\text{\tiny \it B}}_{k} + {\rm H.C}\Big)\,,\,{\rm Reg. }~B\,,\\\displaystyle \Big(e^{-ik{v}_\text{\tiny \it C}}\hat{a}^{\text{\tiny \it C}}_{-k} + {\rm H.C}\Big) +\Big(e^{+ik{u}_\text{\tiny \it C}}\hat{a}^{\text{\tiny \it C}}_{k} + {\rm H.C}\Big)\,,\,{\rm Reg. }~C\,.
\end{dcases}}
where '${\rm H.C.}$' stands for 'hermitian conjugate'. The form in (\ref{eq:mode1}) can be used to define the scalar field $\hat{\phi}$ on the entire expanding FLRW patch, precisely as (\ref{eq:mode}). The crucial point is that in general the Bunch-Davies vacuum as defined by (\ref{eq:vac}) is an entangled combination of states inside and outside the horizon, which leads to an increase in entropy once the hidden states are traced over.

We will first perform the entire calculation for $\hat{\phi}_V$, after which writing the results for $\hat{\phi}_U$ becomes trivial. We can first focus only on the regions $A$ and $B$, since for $\hat{\phi}_V$ only the horizon at $V=0$ is relevant, which we elaborate more below. From (\ref{eq:mode1}) we then get \ea{\hat{\phi}_V=\int_0^\infty\f{dk}{\sqrt{4\pi k}}\Big[e^{-ik{v}_\text{\tiny \it A}}\hat{a}^{\text{\tiny \it A}}_{-k} &+
e^{ik{v}_\text{\tiny \it A}}\hat{a}^{\text{\tiny \it A}\,\dagger}_{-k}\nonumber
\\&\label{eq:modeeq0}+e^{ik{v}_\text{\tiny \it B}}\hat{a}^{\text{\tiny \it B}}_{-k} +e^{-ik{v}_\text{\tiny \it B}}\hat{a}^{\text{\tiny \it B}\,\dagger}_{-k}\Big]\,,}
where the modes defined in the $A$ region are to be understood to vanish in region $B$ and vice versa for the modes in $B$.

The approach we will use was originally presented in \cite{Unruh:1976db} and is based on the fact that any linear combination of positive modes defines the same vacuum as a single positive mode \cite{Birrell:1982ix}. The main constraint is that the Bunch-Davies modes (\ref{eq:mode}) are continuous across the horizon. Making use of (\ref{eq:coors21} -- \ref{eq:coors24}) we can write
\ea{e^{-iv_\text{\tiny \it A} k}&=e^{\f{ik}{H}\ln(-HV)}\,;~~ {\rm Reg. }~A\,,\label{eq:f1} \\
e^{-\f{\pi k}{H}}\big(e^{ikv_\text{\tiny \it B}})^*=e^{-\f{\pi k}{H}}e^{\f{ik}{H}\ln(HV)}&=e^{\f{ik}{H}\ln(-HV)}\,;~~ {\rm Reg. }~B\,,\label{eq:f2}}
where in the last line we chose the complex logarithm to have a branch cut as $\ln(-1)=i\pi$. Because of this choice the sum of (\ref{eq:f1}) and (\ref{eq:f2}) is continuous across the horizon and analytic when $\Im [V]<0$, so it must be expressible as a linear combination of $e^{-ikV}$ i.e the positive frequency Bunch-Davies modes from (\ref{eq:mode}). In a similar fashion starting from $e^{+iv_\text{\tiny \it A} k}$ it is straightforward to find a second continuous and well-behaved linear combination. With such linear combinations we have yet another representation for the scalar field in addition to (\ref{eq:mode}) and (\ref{eq:modeeq0})
\ea{\hat{\phi}_V&=\int_0^\infty\f{dk}{\sqrt{4\pi k}}\f{1}{\sqrt{1-\gamma^2}}\bigg\{\Big[e^{-ikv_\text{\tiny \it A}}+\gamma\big(e^{ikv_\text{\tiny \it B} })^*\Big]\hat{d}^{(1)}_{-k}\nonumber\\&+\Big[\gamma\big(e^{-ikv_\text{\tiny \it A}})^*+e^{ikv_\text{\tiny \it B}}\Big]\hat{d}^{(2)}_{-k}+ {\rm H.C.}\bigg\}\,,\label{eq:modeeq}}
where we have defined $\gamma\equiv e^{-\pi k/H}$ and used (\ref{eq:ipa} -- \ref{eq:ipb}) to get properly normalized modes. An important result may be derived by realizing that the operators $\hat{d}^{(1)}_{-k}$ and $\hat{d}^{(2)}_{-k}$ annihilate the Bunch-Davies vacuum since the modes in the square brackets of (\ref{eq:modeeq}) are continuous and must be linear combinations of the positive frequency Bunch-Davies modes and (\ref{eq:modeeq0}) allows us to express them in terms of $\hat{a}^{\text{\tiny \it A}}_{-k}$ and $\hat{a}^{\text{\tiny \it B}}_{-k}$
\ee{
\begin{cases}\big(\hat{a}^{\text{\tiny \it A}}_{-k}-\gamma\hat{a}^{\text{\tiny \it B}\,\dagger}_{-k}\big)|0_{-k}\rangle&=0\\
\big(\hat{a}^{\text{\tiny \it B}}_{-k}-\gamma\hat{a}^{\text{\tiny \it A}\,\dagger}_{-k}\big)|0_{-k}\rangle&=0\end{cases}
\,.\label{eq:lincomb3}}
The above relations are identical to what is found in the 2-dimensional Unruh effect, and black hole evaporation and imply that  $|0_{-k}\rangle$ is an entangled state in terms the number bases as defined by $\hat{a}^{\text{\tiny \it A}}_{-k}$ and $\hat{a}^{\text{\tiny \it B}}_{-k}$. Following \cite{Fiola:1994ir} the normalized solution to (\ref{eq:lincomb3}) can be written as
\ee{|0_{-k}\rangle=\sqrt{1-\gamma^2}\sum_{n_{-k}=0}^\infty\gamma^{n_{-k}}|n_{-k},A\rangle|n_{-k},B\rangle\,,\label{eq:aniis}}
where $|n_{-k},A\rangle$ and $|n_{-k},B\rangle$ are particle number eigenstates as defined by $\hat{a}^{\text{\tiny \it A}}_{-k}$ and $\hat{a}^{\text{\tiny \it B}}_{-k}$. If as in (\ref{eq:cg}) we trace over the unobservable states the density matrix becomes precisely thermal
\ea{\hat{\rho}_{-{k}}&={\rm Tr}_\text{\it B}\big\{|0_{-k}\rangle\langle0_{-k}|\big\}\nonumber \\&=(1-\gamma^2)\sum_{n_{-k}=0}^\infty\gamma^{2n_{-k}}|n_{-k},A\rangle\langle A, n_{-k}|\,.}

A physical argument can also be used to rule out one of the two possible choices for the branch cut. Had we made the different choice the result would have given an infinite number of produced particles at the ultraviolet limit, which is an unphysical solution as the ultraviolet modes should be indifferent to the global structure of spacetime and experience no particle creation.

As mentioned, only the regions $A$ and $B$ are relevant for $\hat{\phi}_V$. The reason why one may neglect the contribution from region $C$ is apparent from the relations (\ref{eq:coors21} -- \ref{eq:coors24}) and (\ref{eq:mode1}): $\hat{\phi}_V$ has no dependence on $U$ so no mixing of modes is needed in order to obtain analytic behaviour across the horizon $U=0$. Thus, including all regions $A,B$ and $C$ in (\ref{eq:modeeq}) would still give (\ref{eq:aniis}), which we have also explicitly checked.

So far we have only studied $\hat{\phi}_V$ i.e. the particles moving to the left. By using (\ref{eq:coors21} -- \ref{eq:coors24}) and (\ref{eq:mode1}) the calculation involving $\hat{\phi}_U$  proceeds in an identical manner resulting also in a thermal density matrix, but in terms of the right-moving particles $|n_k,A\rangle$.

Putting everything together, the density matrix (\ref{eq:dmc}) obtained by tracing over the states beyond the horizon in the Bunch-Davies vacuum is precisely thermal with the Gibbons-Hawking de Sitter temperature $T=H/(2\pi)$ 
\ee{\hat{\rho}=\prod_{\mathbf{k}}\big(1-e^{- \f{2\pi k}{H}}\big)\sum_{n_{\mathbf{k}}=0}^\infty e^{- \f{2\pi k}{H}n_{\mathbf{k}}}|n_{\mathbf{k}},A\rangle\langle A, n_{\mathbf{k}}|\,.\label{eq:desnm}}

Since in (\ref{eq:desnm}) all states except the ones belonging to region $A$ are neglected we can write the expectation value of the energy-momentum tensor from (\ref{eq:munu}) by expressing $\hat{\phi}$ as the top line from (\ref{eq:mode1}) 
\ee{\langle\hat{T}_{vv}\rangle=\langle\hat{T}_{uu}\rangle=\int_0^\infty\f{dk}{2\pi}\bigg[\f{k}{2}+\f{k}{e^{2\pi k/H}-1}\bigg]\,;~~~\langle\hat{T}_{uv}\rangle=0\,,\label{eq:tuu}}
where for simplicity we have dropped the $A$ labels. The final unrenormalized expression in the FLRW coordinates (\ref{eq:FLRW2}) can be obtained by using the tensor transformation law $T_{\mu\nu}=\f{\partial x^{\tilde{\alpha}}}{\partial x^\mu}\f{\partial x^{\tilde{\beta}}}{\partial x^\nu}T_{\tilde{\alpha}\tilde{\beta}}$ with (\ref{eq:coors21} -- \ref{eq:coors24}) and (\ref{eq:LC0}). This gives
\ea{\label{eq:t000}\langle\hat{T}_{00}\rangle=\langle\hat{T}_{ii}\rangle/a^2&=\bigg[\f{1}{(1-H\ubar{x})^{2}}+\f{1}{(1+H\ubar{x})^{2}}\bigg]\nonumber\\&\times\int_0^\infty\f{dk}{2\pi}\bigg[\f{k}{2}+\f{k}{e^{2\pi k/H}-1}\bigg]\,,}and\ea{
\langle\hat{T}_{i0}\rangle/a&=\bigg[\f{1}{(1-H\ubar{x})^{2}}-\f{1}{(1+H\ubar{x})^{2}}\bigg]\nonumber\\&\times\int_0^\infty\f{dk}{2\pi}\bigg[\f{k}{2}+\f{k}{e^{2\pi k/H}-1}\bigg]\,.\label{eq:t00i}}

As the discussion after equation (\ref{eq:cg}) addressed, (\ref{eq:t000}) and (\ref{eq:t00i}) have divergent behaviour on the horizons $\ubar{x}=\pm H^{-1}$. This is distinct to the usual ultraviolet divergences encountered in quantum field theory, which are also present in (\ref{eq:t000}) and (\ref{eq:t00i}) as the divergent integrals. For our purposes the relevant limit of being close to the origin is obtained with an expansion in terms of $H\ubar{x}$ giving
\ea{\langle\hat{T}_{00}\rangle=\langle\hat{T}_{ii}\rangle/a^2&=\int\f{d\mathbf{k}}{2\pi}\bigg[\f{k}{2}+\f{k}{e^{2\pi k/H}-1}\bigg]+\mathcal{O}(H\ubar{x})^2\\
\langle\hat{T}_{i0}\rangle/a&=H\ubar{x}\big[\langle\hat{T}_{00}\rangle+\langle\hat{T}_{ii}\rangle/a^2 \big]\label{eq:t00ii}+\mathcal{O}(H\ubar{x})^3\,.}

The last step in the calculation is renormalization. When we neglect the $\mathcal{O}(H\ubar{x})$ contributions i.e. study only the region far from the horizon the result is precisely homogeneous and isotropic for which the counter terms can be found by calculating the energy-momentum tensor as an expansion in terms of derivatives the scale factor. This is the adiabatic subtraction technique \cite{Parker:1974qw1, Bunch:1980vc,Parker:1974qw}, with which the 2-dimensional counter terms were first calculated in \cite{Bunch:1978gb} giving coinciding results to \cite{Davies:1976ei}. This technique gives the counter terms for the energy and pressure components as formally divergent integrals
\ee{\delta T_{00}=\int\f{d\mathbf{k}}{2\pi}\f{k}{2}+\f{H^2}{24\pi}\,;\qquad
\delta T_{ii}/a^2=\int\f{d\mathbf{k}}{2\pi}\f{k}{2}-\f{H^2}{24\pi}\,. }
Note that the apparent divergence in the flux contribution (\ref{eq:t00ii}) is an artefact of our use of non-regulated integrals. When dimensionally regulated the sum of the energy and  pressure density divergences cancels, also for massive particles. Physically one can understand this from the requirement that Minkowski space must be stable under back reaction. This issue is discussed more in section \ref{sec:br}, see also the equation (\ref{eq:rpm}). For now we can simply neglect the divergences in (\ref{eq:t00ii}) or following \cite{Markkanen:2016vrp} formally derive the flux counter terms by demanding covariant conservation of $\delta T_{\mu\nu}$.  
The renormalized energy-momentum is then
\ea{\label{eq:00}\rho_{m}&=\int\f{d\mathbf{k}}{2\pi}\f{k}{e^{2\pi k/H}-1}-\f{H^2}{24\pi}+\mathcal{O}(H\ubar{x})^2\,,\\p_m&=\label{eq:ii}\int\f{d\mathbf{k}}{2\pi}\f{k}{e^{2\pi k/H}-1}+\f{H^2}{24\pi}+\mathcal{O}(H\ubar{x})^2\,,\\{T}_{i0}/a\equiv f_m&=H\ubar{x}\big(\rho_m+p_m\big)+\mathcal{O}(H\ubar{x})^3\,.\label{eq:flux}} 
We can clearly see that far away from the horizon the energy-momentum describes a homogeneous and isotropic distribution of thermal particles. It can be checked to be covariantly conserved and to have the usual conformal anomaly \cite{Birrell:1982ix}, although in de Sitter space the conformal anomaly is not important and one may cancel the $\pm \f{H^2}{24\pi}$ contributions by a redefinition of the cosmological constant. Contrary to what one usually encounters in cosmology, the energy density in (\ref{eq:00}) is constant despite the fact that space is expanding and generically diluting any existing particle density. This is explained by the additional term (\ref{eq:flux}) representing a continuous incoming flux of particles, which replenishes the energy lost by dilution. This naturally raises an important question concerning the source of the flux, which will be addressed in section \ref{sec:br} where we write down solutions that are consistent in terms of semi-classical back reaction.

In the language of section \ref{sec:gen} the $\pm H^2/(24\pi)$ terms are state independent contributions resulting from renormalization and the integral over the thermal distribution is the  state dependent contribution $\rho^S_m$. The results (\ref{eq:00} -- \ref{eq:flux}) can be seen to be in agreement with the arguments of section \ref{sec:gen}, in particular the 2-dimensional version of (\ref{eq:rpq}) when taking into account of the modifications arising from our choice of not to implement dimensional regularization.

Before ending this subsection we comment on a technical detail regarding the divergences generated in the coarse grained state. A general -- or certainly a desirable -- feature of quantum field theory is the universality of the generated divergences and renormalization. For a quantum field on a de Sitter background, one should be able to absorb all divergences in the redefinition of the cosmological constant, preferably also in the coarse grained state (\ref{eq:desnm}). But from (\ref{eq:t000}) and (\ref{eq:t00i}) we can see  that this does not hold due to the $\ubar{x}$-dependence of the generated divergences, likely related to coarse graining and the additional divergence when approaching the horizon. 
We emphasize however that even if in a carefully defined coarse graining divergences $\propto\ubar{x}$ are not generated, the renormalized result would coincide with (\ref{eq:00} -- \ref{eq:flux}), because up to the accuracy we are interested in all the needed counter terms could be derived via adiabatic subtraction, which satisfies $\delta T_{00}=-\delta T_{ii}/a^2$ \cite{Markkanen:2016aes}. 

\subsection{Four dimensions}
\label{sec:4}
The calculation in four dimensions proceeds in principle precisely as the 2-dimensional derivation of the previous subsection. The main differences are that the solutions in four dimensions are analytically more involved and that there is only one horizon. 
Quantization of a scalar field in the static de Sitter patch has been studied in \cite{Higuchi:1986ww,Polarski:1990tr,loh} to which we refer the reader for more details. Quite interestingly, although the line element for a Schwarzschild black hole and de Sitter space in static coordinates are very similar, the latter has an analytic solution for the modes while the former does not.

We begin by writing useful coordinate transformations between the 4-dimensional FLRW coordinates and static coordinates of section \ref{sec:coors}, in region $A$
\ee{\begin{cases}e^{Ht}\!\!\!\!&{=}~e^{H\ubar{t}_\text{\tiny \it A}}\sqrt{1-(H \ubar{r}_\text{\tiny \it A})^2}\\
r\!\!&{=}~e^{-Ht}\ubar{r}_\text{\tiny \it A}\end{cases}\,,\,{\rm Reg. }~A\,,\label{eq:coors33}}and region $B$ \ee{\begin{cases}e^{Ht}\!\!\!\!&{=}~e^{H\ubar{t}_\text{\tiny \it B}}\sqrt{(H \ubar{r}_\text{\tiny \it B})^2-1}\\
r\!\!&{=}~e^{-Ht}\ubar{r}_\text{\tiny \it B}\end{cases}\,,\,{\rm Reg. }~B\,. \label{eq:coors3}}
The tortoise coordinates and light-cone coordinates can be obtained trivially from the 2-dimensional results (\ref{eq:LC0}), (\ref{eq:tort}) and (\ref{eq:coors21} -- \ref{eq:coors24}) with the replacements ${x}\rightarrow {r}$ and  $\ubar{x}\rightarrow \ubar{r}$.

Due to spherical symmetry and time-independence of the metric (\ref{eq:sta}) in the coordinates (\ref{eq:staA}) we introduce a similar ansatz to the equation of motion (\ref{eq:eom}) as in the spherical form of the FLRW metric in (\ref{eq:ans4})
\ee{\psi_{\ell m k}^\text{\tiny \it  A}=f^\text{\tiny \it  A}_{\ell k}(\ubar{r}_\text{\tiny \it  A})Y^m_\ell e^{-ik\ubar{t}_\text{\tiny \it  A}}\label{eq:ans}\,,}
leading to the radial equation
\ea{\bigg\{&\ubar{r}_\text{\tiny \it  A}^{-2}\f{\partial}{\partial{\ubar{r}_\text{\tiny \it  A}}}\Big(\ubar{r}_\text{\tiny \it  A}^2\big[1-(H\ubar{r}_\text{\tiny \it  A})^2\big]\f{\partial}{\partial{\ubar{r}_\text{\tiny \it  A}}}\Big)+\f{k^2}{1-(H\ubar{r}_\text{\tiny \it A})^2} \nonumber \\&-\f{\ell(\ell+1)}{\ubar{r}_\text{\tiny \it  A}^2} -2H^2\bigg\}f^\text{\tiny \it  A}_{\ell k}(\ubar{r}_\text{\tiny \it  A})=0\,,\label{eq:ansa}}
where we have used $\xi_4 R = 2H^2$. With a suitable ansatz the above may be reduced to a hypergeometric equation and has the solution in terms of the Gaussian hypergeometric function
\ea{&f^\text{\tiny \it  A}_{\ell k}(\ubar{r}_\text{\tiny \it  A})=D^\text{\tiny \it  A}_{\ell k}(H\ubar{r}_\text{\tiny \it  A})^\ell\big[1-(H\ubar{r}_\text{\tiny \it  A})^2\big]^{\f{ik}{2H}}\nonumber \\ &\times{\,}_2F_1\bigg[\f{\ell}{2} +\f{ik}{2H} +\f{1}{2},\f{\ell}{2} +\f{ik}{2H} +{1};\ell +\f{3}{2};(H\ubar{r}_\text{\tiny \it  A})^2\bigg]\,,\label{eq:f}}
where $D^\text{\tiny \it  A}_{\ell k}$ is a normalization constant. With the help of (\ref{eq:coors33}) and (\ref{eq:coors3}) we can show that 
\ee{\partial_{\ubar{t}_\text{\tiny \it  A}}=e^{-Ht}\big[\partial_\eta-H\ubar{r}_\text{\tiny \it  A}\partial_{r}\big]\,,}
so inside the horizon in the static coordinates we can use $\partial_{\ubar{t}_\text{\tiny \it  A}}$ for defining positive frequency modes and the inner product via (\ref{eq:norma}), which reads
\ee{(\phi_1,\phi_2\big)_\text{\tiny \it  A}=-i\int d\Omega\int_0^{1/H}d\ubar{r}_\text{\tiny \it  A}\f{\ubar{r}_\text{\tiny \it  A}^2}{1-(H\ubar{r}_\text{\tiny \it  A})^2}\phi_1^{\phantom{*}}\overset{\leftrightarrow}{\nabla}_{\ubar{t}_\text{\tiny \it A}}\phi^*_2\,.
}
Using (\ref{eq:ans}) and (\ref{eq:f}) in the above allows one to solve for the normalization constant $D_{\ell k}$. For details we refer the reader to \cite{Higuchi:1986ww,Polarski:1990tr} and appendix \ref{sec:B}, but here we simply write the result
\ea{\f{1}{D_{\ell k}^{_\text{\tiny \it  A}}}&=\f{\sqrt{4\pi k}}{H}\!{\,}_2F_1\bigg[\f{\ell}{2} +\f{ik}{2H} +\f{1}{2},\f{\ell}{2} +\f{ik}{2H} +{1};\ell +\f{3}{2};1\bigg]\nonumber \\&=\f{\sqrt{4\pi k}\Gamma\big[\ell +\f{3}{2}\big]\Gamma\big[\f{-ik}{H}\big]}{H\Gamma\big[\f{1}{2}\big(\ell - \f{ik}{H} +1\big)\big]\Gamma\big[\f{1}{2}\big(\ell - \f{ik}{H} +2\big)\big]}\,,\label{eq:ncost}}
up to factors of modulus one.

Having derived the solutions in the static patch inside the horizon (\ref{eq:staA}),  the solutions in coordinates covering the outside of the horizon (\ref{eq:staB}) follow trivially from the fact that the line element and thus the radial equation (\ref{eq:ansa}) have identical form. Again however, we must carefully determine the correct normalization for the positive frequency mode beyond the horizon. With (\ref{eq:coors33}) and (\ref{eq:coors3}) we can show 
\ee{\partial_{\ubar{r}_\text{\tiny \it  B}}=\f{e^{-Ht}}{(H\ubar{r}_\text{\tiny \it  B})^2-1}\big[H\ubar{r}_\text{\tiny \it  B}\partial_\eta-\partial_{r}\big]\,,\label{eq:vcet}}
which, similarly to the 2-dimensional derivation, implies that $\ubar{r}_\text{\tiny \it  B}$ plays the role of time. The inner product in the $B$ region then becomes
\ee{(\phi_1,\phi_2\big)_\text{\tiny \it B}=-i\int d\Omega\int_{-\infty}^{\infty}d\ubar{t}^{\phantom{2}}_\text{\tiny \it  B}\,{\ubar{r}_\text{\tiny \it  B}^2}\big[(H\ubar{r}_\text{\tiny \it  B})^2-1\big]\phi_1^{\phantom{*}}\overset{\leftrightarrow}{\nabla}_{\ubar{r}_\text{\tiny \it B}}\phi^*_2\,,
\label{eq:ippe}}
Since the inner product (\ref{eq:ippe}) is independent of the choice of hypersurface we can evaluate it at the limit $\ubar{r}_\text{\tiny \it  B}\rightarrow 1/H$. It is then a simple matter of using (\ref{eq:ncost}) to show that
\ee{\psi_{\ell m k}^\text{\tiny \it  B}=f_{\ell k}^\text{\tiny \it  B}(\ubar{r}_\text{\tiny \it  B})\big(Y^m_\ell\big)^* e^{ik\ubar{t}_\text{\tiny \it  B}}\label{eq:ans2}\,,}
in region $B$ is the correctly normalized positive frequency mode provided that
\ea{&f^\text{\tiny \it  B}_{\ell k}(\ubar{r}_\text{\tiny \it  B})=D^\text{\tiny \it  B}_{\ell k}(H\ubar{r}_\text{\tiny \it  B})^\ell\big[(H\ubar{r}_\text{\tiny \it  B})^2-1\big]^{-\f{ik}{2H}}\nonumber \\ &\times{\,}_2F_1\bigg[\f{\ell}{2} -\f{ik}{2H} +\f{1}{2},\f{\ell}{2} -\f{ik}{2H} +{1};\ell +\f{3}{2};(H\ubar{r}_\text{\tiny \it  B})^2\bigg]\,,\label{eq:fb}} 
with $D^\text{\tiny \it  B}_{\ell k}=(D^\text{\tiny \it  A}_{\ell k})^*$, see appendix \ref{sec:B}.

In the 4-dimensional expanding FLRW patch the scalar field can then be written in the coordinates (\ref{eq:staA}) and (\ref{eq:staB}) as
\ee{\hat{\phi}=\begin{dcases}\sum_{\ell =0}^\infty\sum_{m=-\ell}^\ell\int_0^\infty dk\Big[\psi_{\ell m k}^\text{\tiny \it  A}\hat{a}^\text{\tiny \it A}_{\ell m k}+{\rm H.C.}\Big]\qquad{\rm Reg. }~A\,,\\\sum_{\ell =0}^\infty\sum_{m=-\ell}^\ell\int_0^\infty dk\Big[\psi_{\ell m k}^\text{\tiny \it  B}\hat{a}^\text{\tiny \it B}_{\ell m k}+{\rm H.C.}\Big]\qquad{\rm Reg. }~B\,.\end{dcases} \label{eq:psta2}}

Despite the analytically involved structure of the modes the arguments we used in the 2-dimensional case apply practically identically in four dimensions. This is due to the simplification that occurs when approaching the horizon. At this limit with the help of (\ref{eq:ncost}) and the tortoise coordinates (\ref{eq:tort}) we get for the mode in region $A$
\ea{\psi_{\ell m k}^\text{\tiny \it  A}&\overset{H\ubar{r}_\text{\tiny \it  A}\rightarrow 1}{\longrightarrow}H\f{\phantom{\big(}Y^m_\ell\phantom{\big)^*}}{\sqrt{4\pi k}}\Big\{\big[\cosh(H\ubar{r}^*_\text{\tiny \it  A})\big]^{-2}\Big\}^{+\f{ik}{2H}}e^{-ik\ubar{t}_\text{\tiny \it  A}} \nonumber\\&\quad\sim \f{\phantom{\big(}Y^m_\ell\phantom{\big)^*}}{\sqrt{4\pi k}\ubar{r}_\text{\tiny \it  A}}e^{-ikv_\text{\tiny \it A}}\,,\label{limab}} and for the mode in region $B$\ea{\psi_{\ell m k}^\text{\tiny \it  B}&\overset{H\ubar{r}_\text{\tiny \it  B}\rightarrow 1}{\longrightarrow}H\f{\big(Y^m_\ell\big)^*}{\sqrt{4\pi k}}\Big\{\big[\sinh(H\ubar{r}^*_\text{\tiny \it  B})\big]^{-2}\,\Big\}^{-\f{ik}{2H}}e^{+ik\ubar{t}_\text{\tiny \it  B}} \nonumber\\&\quad\sim \f{{\big(}Y^m_\ell{\big)^*}}{\sqrt{4\pi k}\ubar{r}_\text{\tiny \it  B}}e^{+ikv_\text{\tiny \it  B}}\,,\label{limbb}}
where $\sim$ denotes an equality up to constant factors of modulus one. Comparing the above to (\ref{eq:mode1}) reveals that the discontinuity is precisely of the same form as in two dimensions, leading to two non-trivial linear combinations that are continuous across the horizon 
\ea{\psi_{\ell m k}^\text{\tiny \it  A}+\gamma(\psi_{\ell m k}^\text{\tiny \it  B})^*\qquad \text{and}\qquad\gamma(\psi_{\ell m k}^\text{\tiny \it  A})^*+\psi_{\ell m k}^\text{\tiny \it  B}\,,\label{eq:modeeq2} }
where again $\gamma\equiv e^{-\pi k/H}$. Expressing the scalar field in terms of two representations, as was done in two dimensions in (\ref{eq:modeeq}), one may reproduce the steps of the 2-dimensional derivation and deduce the thermality of the 4-dimensional coarse grained density matrix
\ee{\hat{\rho}=\prod_{\ell m k}\big(1-e^{-\f{2\pi k}{H} }\big)\sum_{n_{\ell m k}=0}^\infty e^{- \f{2\pi k}{H}n_{\ell m k}}|n_{\ell m k},A\rangle\langle A, n_{\ell m k}|\,.\label{eq:desnm2}}

The energy density in four dimensions is of course more difficult to write in a clear form than the 2-dimensional result due to the presence of the hypergeometric functions in the mode solution (\ref{eq:f}). However for our purposes only the limiting case of a being far from the horizon is relevant, for which the static line element (\ref{eq:sta}) coincides with flat space up to small terms $\mathcal{O}(H\ubar{x})$. This and the fact that our theory is conformal imply that up to $\mathcal{O}(H\ubar{x})$ the energy density in the coarse grained state (\ref{eq:desnm2}) should coincide with that of a black-body with $T=H/(2\pi)$. In appendix \ref{sec:A} we verify this assertion with an explicit calculation. 

Tracelessness of the unrenormalized energy-momentum tensor immediately fixes the ratio of the energy and pressure densities, again dropping the $A$ labels as irrelevant
\ea{\langle\hat{T}_{\,\ubar{\!0}\,\ubar{\!0}} \rangle=3\langle\hat{T}_{\ubar{i}\,\ubar{i}}\rangle= \int\f{d^3\mathbf{k}}{(2\pi)^3}\bigg[\f{k}{2}+\f{k}{e^{2\pi k/H}-1}\bigg]+\mathcal{O}(H\ubar{x})^2
\, ,\label{eq:t00}}
where the off-diagonal components cancel due to the lack of flux and shear in a static spherically symmetric case. From the results expressed in the static coordinates in (\ref{eq:t00}) we already see a very important outcome from coarse graining the energy-momentum tensor with respect to states beyond the horizon: the sum of the pressure- and energy-density does not vanish, a relation which is satisfied by the '$00$' and '$ii$' components of Einstein tensor in the static coordinates at the centre of the Hubble sphere. This implies that there is non-trivial back reaction since Einstein's equation with (\ref{eq:t00}) as the source is not solved by the static line element describing de Sitter space (\ref{eq:sta}). We can conclude that coarse graining such that only observable states are left leads to a violation of de Sitter invariance.

 Following the 2-dimensional procedure of the previous subsection, the usual tensor transformations with the help of (\ref{eq:coors33}) and (\ref{eq:coors3}) allow us to write the result in the FLRW form, which we can renormalize by using the 4-dimensional adiabatic counter terms that can be found for example in \cite{Markkanen:2016jhg}. The final result is
\ea{\label{eq:ft0}\rho_m&=\int\f{d^3\mathbf{k}}{(2\pi)^3}\f{k}{e^{2\pi k/H}-1}+\f{H^4}{960\pi^2}\,,\\\label{eq:ft1}p_m&=\f{1}{3}\int\f{d^3\mathbf{k}}{(2\pi)^3}\f{k}{e^{2\pi k/H}-1}-\f{H^4}{960\pi^2}\,,\\f_{m,\,i}&=H\ubar{x}^i\big(\rho_m+p_m\big)\,\label{eq:Tmunu},}
where $f_{m,\,i}\equiv T_{i0}/a$ and we have neglected terms of $\mathcal{O}(H\ubar{x})^2$. So quite naturally, the 4-dimensional result has exactly the same thermal characteristics to the 2-dimensional one in (\ref{eq:00} -- \ref{eq:flux}). Far away from the horizon (\ref{eq:ft0} -- \ref{eq:Tmunu}) is homogeneous and isotropic with
\ee{\rho_m+p_m=\f{4}{3}\rho^S_m=\f{4}{3}\int\f{d^3\mathbf{k}}{(2\pi)^3}\f{k}{e^{2\pi k/H}-1}=\f{H^4}{360\pi^2}\,, \label{eq:evol}}
and again the terms $\pm \f{H^4}{960\pi^2}$ responsible for the conformal anomaly are irrelevant and could have been removed with a redefinition of $\Lambda$. The above can be seen to be in agreement with the discussion of section \ref{sec:gen}, specifically the right-hand side of equation (\ref{eq:import}).

For clarity we summarize the arguments of this section here once more: in de Sitter space as described by the expanding FLRW coordinates (\ref{eq:FLRW3}) initialized to the Bunch-Davies vacuum the energy-momentum of a quantum field has a thermal character when in the density matrix one includes only the observable states inside the horizon. The energy density inside the horizon is maintained at a constant temperature by a continuous flux of radiation incoming from the horizon that precisely cancels the dilution from expansion. This results in $\rho_m+p_m\neq0 $, which is independent of the details of renormalization and the conformal anomaly due to the symmetries of the counter terms in de Sitter space. Thus even at the limit when the distance to the horizon is very large and the result is isotropic and homogeneous the sum of the energy and pressure densities does not cancel and because of this the dynamical Friedmann equation (\ref{eq:Hev}) then implies that a strictly constant Hubble rate $H$ is not a consistent solution. This is also visible in the result given in the static coordinates (\ref{eq:t00}), which does not solve Einstein's equation if the background is assumed to be strictly de Sitter.
 
\section{Self-consistent back reaction}
\label{sec:br}
If, as the arguments of the previous section imply, de Sitter space is affected by back reaction in the prescription we have chosen, this naturally leads one to investigate how precisely is the strict de Sitter solution modified. Ultimately, this is determined by the semi-classical Einstein equation.

Taking the limit of begin close to the center of the Hubble sphere, which is the same as assuming that the horizon is far away, the equations (\ref{eq:ft0} -- \ref{eq:Tmunu}) correspond to a homogeneous and an isotropic solution. One would then expect that at this limit also the back reaction is homogeneous and isotropic parametrizable with a FLRW line element. In fact strictly speaking, we can only consistently study back reaction if the homogeneous and isotropic approximation holds, since the calculation of the previous section was made by assuming the FLRW line element. In this case the quantum corrected Hubble rate can be self-consistently solved by using
\ee{\rho_m=\f{H^4}{480\pi^2}\,;\qquad p_m=\f{1}{3}\f{H^4}{480\pi^2}\,,\label{eq:Tmunu1}}
where in the above for simplicity we have absorbed the $\pm {H^4}/({960\pi^2})$ contributions in (\ref{eq:ft0}) and (\ref{eq:ft1}) into the (re)definition of the cosmological constant, making the distinction between $\rho^S_m$ and $\rho_m$ irrelevant and left the $\mathcal{O}(H\ubar{x})$ notation as implicit. However, before proceeding we must address a crucially important implication of having a constant energy and pressure density in a spacetime described by a FLRW line element: (\ref{eq:Tmunu1}) are not consistent with covariant conservation \ee{\dot{\rho}+3H(\rho+p)=0\,,\label{eq:covcons}}
where $\rho=\rho_m+\rho_\Lambda$ and similarly for $p$. This is expected, since we have not included the effect of the flux (\ref{eq:Tmunu}), which continuously injects the system with more energy. 

As argued in \cite{Hawking:1974sw} for the analogous black hole case a flux can be seen to imply a change in the size of horizon: a positive flux coming from the horizon is equivalent to a negative flux going into the horizon. Note that when the cosmological horizon or $\rho_\Lambda$ absorbs negative energy the horizon radius will grow, where as precisely the opposite relation holds for the horizon and mass of a black hole. 

If we make the assumption that the rate of change of the vacuum energy equals the energy injected by the flux (\ref{eq:Tmunu}) we can satisfy (\ref{eq:covcons}) while matching the right-hand side of the dynamical Friedmann equation (\ref{eq:Hev}) with  (\ref{eq:evol}), by re-writing the Friedmann equations (\ref{eq:e}) with dynamical vacuum energy, $\rho_\Lambda\longrightarrow\tilde{\rho}_\Lambda$, for which
\ea{\tilde{\rho}_\Lambda=-\tilde{p}_\Lambda&=\rho_\Lambda-3tH(\rho_m+p_m)=\Lambda M_{\rm pl}^2-t\f{H^5}{120\pi^2}\nonumber \\\Rightarrow~~ \dot{\tilde{\rho}}_\Lambda&=-3H(\rho_m+p_m)
\,,\label{eq:labe}}
where small terms of $\mathcal{O}(\dot{H})$ are beyond our approximation and are neglected. The dynamical Friedmann equation (\ref{eq:Hev}) then allows us to solve for the evolution of $H$ 
\ee{-2\dot{H}M_{\rm pl}^2=\f{H^4}{360\pi^2} \quad\Rightarrow\quad H=\f{H_0}{\Big(\f{H_0^3}{240\pi^2M_{\rm pl}^2}t+1 \Big)^{1/3}}\,.\label{eq:Hev2}}
The crucial observation is that it is impossible for the de Sitter approximation to hold for an arbitrarily long time: after the time scale \ee{t_{1/2}=1680\pi^2\bigg(\f{M_{\rm pl}}{H_0}\bigg)^2H_0^{-1}\,,\label{eq:HL}} 
which is defined as the half-life of the Hubble rate, the system seizes to be de Sitter and we can conclude that the cumulative effect of the quantum back reaction has become significant enough to dominate over the classical solution. The time scale for the destabilization of de Sitter $t\sim M_{\rm pl}^2/H_0^3$ was similarly obtained for a fully quantized model in \cite{Dvali:2017eba}, where it is also argued that after this time the full quantum evolution must depart completely from the classical one, in agreement with our analysis\footnote{We thank the authors of \cite{Dvali:2017eba} for clarifying this issue.}. Furthermore, it also follows from the results of section 10.4 of \cite{Padmanabhan:2002ji}.

The time scale (\ref{eq:HL}) is quite large, at least in the obvious cosmological applications: for inflation with the maximum scale allowed by the non-observation of tensor modes $H_0\sim 10^{14}$GeV \cite{Array:2015xqh} gives $t_{1/2}\sim 10^{13}H_0^{-1}$, which corresponds to $10^{13}$ $e$-folds of inflation. The breakdown of the Dark Energy dominated late time de Sitter phase can be estimated by using the current Hubble rate $H_0\sim 10^{-42}$GeV \cite{Ade:2015xua} giving $t_{1/2}\sim 10^{125}H_0^{-1}$, where $H_0^{-1}$ corresponds to the age of the Universe.

The observation that in an expanding, homogeneous and isotropic space a non-diluting particle density necessitates a decaying vacuum energy was already made in \cite{Ozer:1985wr,Freese:1986dd} and has since been studied in \cite{clif,Sola:2016zeg,Lima:1995ea,Alcaniz:2005dg,Prigogine:1989zz,Wands:2012vg,Wang:2015wga}. 


The use of the results of the previous section, which were calculated on a fixed background, is a good approximation only when the modification from back reaction is very small. 
Since without back reaction the result is strict de Sitter space this translates as demanding validity of the adiabatic limit i.e. $H$ should change very gradually. From (\ref{eq:Hev2}) we see this to be true, $-\dot{H}/H^2\sim H^2/M_{\rm pl}^2\ll1$ implying that the quantum modes as well as the right hand side of the Friedmann equations can be calculated in the approximation where the derivatives of $H$ are neglected. One may furthermore check the robustness of the cosmological event horizon and our coarse graining prescription under back reaction: from (\ref{eq:Hev2}) one gets the scale factor \ee{a(t)\propto \exp\bigg\{\f{360\pi^2 M_{\rm pl}^2}{H_0^2}\bigg(\f{H_0^3 }{240\pi^2M_{\rm pl}^2}t+1\bigg)^{2/3}\bigg\}\,,} with which the cosmological event horizon (\ref{eq:hor}) in the presence of back reaction reads
\ee{a(t)\int_t^\infty\f{dt'}{a(t')}=H^{-1}\Big[1+\mathcal{O}\big(H/M_{\rm pl}\big)^2\Big]\,,}
so to a very good approximation the event horizon tracks $1/H$, as required.

The problem with vacuum energy changing its value is that classically it is proportional to a constant parameter $\Lambda$ in the gravitational Lagrangian (\ref{eq:EH}) and it is not obvious how a parameter characterising different de Sitter configurations can change dynamically. In a quantized theory however, the situation changes completely since the zero-point energy and pressure, which all particles possess, satisfy the same equation of state as the contribution resulting from the cosmological constant $\Lambda$. For example, for a massive scalar field when dimensionally regularizing the sum of the zero-point energy and pressure contributions one has 
\ea{\int \f{d^{n-1 }\mathbf{k}}{(2\pi)^{n-1}}\f{\sqrt{\mathbf{k}^2+m^2}}{2}&+\int \f{d^{n-1 }\mathbf{k}}{(2\pi)^{n-1}}\f{\mathbf{k}^2}{2(n-1)\sqrt{\mathbf{k}^2+m^2}}\nonumber \\&=0\,.\label{eq:rpm}}
Unlike in flat space, the zero-point terms are gravitationally significant which is the key issue behind the cosmological constant problem and was first discussed in \cite{Zeldovich1967}. Taking this idea further, if all quantum fields can contribute to the vacuum energy which in turn couples to gravity, it seems natural to assume that in curved space the amount vacuum energy a particular field is responsible for is not fixed, but a dynamical quantity much like the field itself. With this in mind we propose the following physical picture of the continuous particle creation process required by (\ref{eq:Tmunu1}): when a particle pair is created it leaves behind a hole of negative vacuum energy in order not to violate covariant conservation. This causes the overall vacuum energy density to decrease, analogously to the interpretation of black hole evaporation where holes of negative energy fall into the black hole causing it to lose mass \cite{Hawking:1974sw}.

As we discuss in the next section, the proposal that vacuum energy is dynamical also has a deep connection with the thermodynamic interpretation of de Sitter space, which gives it a more solid footing.
\section{Derivation from horizon thermodynamics}

Much like for a black hole, it is expected that the first law of thermodynamics \ee{dU=TdS-PdV\,,\label{eq:FL}} 
can be expressed as a relation connecting internal energy $U$, the horizon and the pressure $P$ of de Sitter space \cite{Gibbons:1977mu,Padmanabhan:2002sha}. However, if the de Sitter solution is assumed to be determined only by the cosmological constant term $\Lambda$, a parameter of the Einstein-Hilbert Lagrangian, the change in internal energy $dU$ requires the problematic concept of varying $\Lambda$ \cite{Spradlin:2001pw,Padmanabhan:2002ji}. If however, we adopt the proposal that semi-classical back reaction is sourced by the coarse grained energy-momentum tensor as discussed in \ref{sec:tr}, the derivation of the two previous sections imply that the vacuum energy becomes a dynamical quantity due to the inevitable contribution of quantum fields 
and this issue is evaded. In fact quite remarkably, allowing the vacuum energy to vary and by using the standard concepts of horizon thermodynamics we can derive the results (\ref{eq:Hev2}) and (\ref{eq:labe}) in a mere few lines. 

In general a spacetime horizon contains entropy proportional to its area
\ee{S=\f{A}{4 G}\,,\label{eq:HE}} which in de Sitter space is given by the de Sitter horizon $A=4\pi H^{-2}$ \cite{Gibbons:1977mu}. We can then define the internal energy contained inside the Hubble sphere as $\rho_\Lambda V=\rho_\Lambda\f{4\pi}{3} H^{-3}$, where $\rho_\Lambda$ is now the dynamical quantity we denoted with $\tilde{\rho}_\Lambda$ in the previous section. In the case of a cosmological horizon there is however an important subtlety giving rise to a few additional minus signs: the change in volume and internal energy in (\ref{eq:FL}) refer to the region that is hidden from the observer, the space beyond the cosmological horizon. Energy lost from the Hubble sphere will in fact be gained by the degrees of freedom beyond the horizon and similarly when the horizon radius increases the volume of the hidden region decreases:
\ee{dU =-d \bigg(\rho_\Lambda \f{4\pi}{3H^3} \bigg)\,;\qquad PdV=-p_{\Lambda}d\bigg( \f{4\pi}{3H^3}\bigg)\,.\label{eq:HE2}}

When the temperature is given by the Gibbons-Hawking relation $T=H/(2\pi)$ and by using the de Sitter equation of state $\rho_\Lambda+p_\Lambda=0$, from the first law (\ref{eq:FL}) with (\ref{eq:HE}) and (\ref{eq:HE2})  we straightforwardly get a relation between the Hubble rate and the change in vacuum energy, 
\ea{-d \bigg(\rho_\Lambda \f{4\pi}{3H^3} \bigg)&=\f{H}{8\pi G}d\bigg(\f{4\pi}{ H^2}\bigg)+p_{\Lambda}d\bigg( \f{4\pi}{3H^3}\bigg)\nonumber} \ea{\Leftrightarrow\quad 2\dot{H}M^2_{\rm pl}&=\f{\dot{\rho}_\Lambda}{3H}\, .\label{eq:ThE}}
If $\rho_\Lambda$ was given by the potential of a scalar field one may recognize (\ref{eq:ThE}) as one of the slow-roll equations used in the inflationary framework \cite{Lyth:2009zz}, which we have here derived by making no reference to Einstein's equation. Then we assume that in addition to vacuum energy the theory contains a massless scalar field with the energy density $\rho_m$. If the de Sitter horizon is a thermodynamic object with the temperature $T=H/(2\pi)$, when given enough time we can expect the cavity enclosed by the horizon to contain a thermal distribution of particles. In the static coordinates (\ref{eq:sta}) this gives the "hot tin can" description of de Sitter space \cite{Frolov:2002va,Kaloper:2002cs,Susskind:2003kw}. However, as we showed in sections \ref{eq:sec3} and \ref{sec:br} the static line element is not a solution of the Einstein equation when the effect of the horizon is included in the quantum averaging leading to a thermal energy-momentum tensor. Hence we must use coordinates that can accomodate also non-de Sitter solutions such as the cosmologically relevant expanding de Sitter patch as described by the FLRW coordinates (\ref{eq:FLRW3}). In these coordinates the tin can picture must be generalized to account for continuous energy loss due to the expansion of space. Simply put, in an expanding space the tin leaks. More concretely, the expansion of space will lead the energy density of the massless particles to dilute as $\propto a^{-4}$, which is a purely geometric statement. We will denote the loss of energy density per unit time from dilution with $\Delta$, which here has the expression
\ee{\Delta=-4H\rho_m\,.\label{eq:dliu}}
If the cosmological horizon maintains thermal equilibrium with an otherwise constantly diluting and thus cooling energy density, an equal amount of heat must flow from the horizon "in" to the bulk that is lost "out" by dilution
\ee{
\dot{\rho}_\Lambda=\Delta\,,\label{eq:out}}
where we have neglected small terms of $\mathcal{O}(\dot{H})$. When the above is inserted in (\ref{eq:ThE}) and with (\ref{eq:dliu}) one gets
\ee{2\dot{H}M^2_{\rm pl}=
-\f{4}{3}\rho_m\,.\label{eq:end}}
For a thermal  $\rho_m$ with the temperature  $T=H/(2\pi)$ equation (\ref{eq:end}) precisely coincides with (\ref{eq:Hev2}) and furthermore the change in vacuum energy (\ref{eq:out}) with (\ref{eq:dliu}) agrees with (\ref{eq:labe}).

Assuming that the thermodynamic features persist even when spacetime has evolved away from de Sitter, as long as the horizon has heat it will continue to radiate and lose energy by dilution and thus to grow without bound. In this case the ultimate fate of the Universe would not be an eternal de Sitter space with finite entropy, but an asymptotically flat spacetime with no temperature, an infinitely large horizon and hence infinite entropy.

\section{Summary and Conclusions}
In this work we have studied the stability of de Sitter space in the semi-classical approach for a model with a non-interacting conformally coupled scalar field and a cosmological constant i.e. vacuum energy. Back reaction was derived in a prescription where the expectation values sourcing the semi-classical Einstein equation were calculated via a coarse grained density matrix containing only states that are observable to a local observer. For the chosen initial condition of the Bunch-Davies vacuum this prescription translates as neglecting all degrees of freedom located beyond the cosmological event horizon. As we have shown via a detailed argument, in our approach de Sitter space is not stable and in agreement with \cite{Padmanabhan:2002ji} (section 10.4) but in disagreement with \cite{Gibbons:1977mu}. 

Coarse graining over unobservable states in the density matrix is made frequent use in various contexts such as the decoherence program and black hole information paradox, but rarely considered in cosmological applications, in particular semi-classical backreaction via the Friedmann equations as done in this work. Our study indicates that loss of information from coarse graining states beyond the horizon leads from the initial Bunch-Davies vacuum, a pure state, to a thermal density matrix and manifestly breaks de Sitter invariance.

Our result also shows that a local observer who is only causally connected to states inside the horizon will in the cosmologically relevant expanding FLRW coordinates view de Sitter space as filled with a thermal energy density with a constant temperature given by the Gibbons-Hawking relation $T=H/(2\pi)$ that is maintained by a continuous incoming flux of energy radiated by the horizon. Without such a flux the expansion of space would dilute and cool the system quickly leading to an empty space.

From the semi-classical Friedmann equations we made the simple but nonetheless important observation that space filled with thermal gas, which in our prescription follows from de Sitter space possessing the cosmological event horizon, is not a solution consistent with having a constant $H$. This follows trivially from the fact that thermal particles do not have the equation of state of vacuum energy and is the key mechanism behind the obtained non-trivial back reaction. This can also be seen in the static coordinates, which are not a solution of Einstein's equation when the energy-momentum tensor describes thermal gas.

By modifying the Friedmann equations to contain gradually decaying vacuum energy we were able to provide a self-consistent solution for the evolution of the Hubble rate. The solution had the behaviour where $H$ remained roughly a constant for a very long time, but eventually after a time scale $\sim M_{\rm pl}^2/H^3$ the system no longer resembled de Sitter space. As a physical picture of the process we proposed that a quantum field in curved space may exchange energy with the vacuum making vacuum energy a dynamical quantity instead of a constant parameter fixed by the Lagrangian. In this interpretation particle creation occurs at the expense of creating a negative vacuum energy contribution. This provides a mechanism allowing the overall vacuum energy to decrease, a very closely analogous picture to black hole evaporation where a black hole loses mass due to a negative energy flux into the horizon. 

Finally, we presented an alternate derivation of the main result by using the techniques of horizon thermodynamics. In the thermodynamic derivation the concept of dynamical vacuum energy proved a crucial ingredient, as it gives a well-defined meaning to the differential of internal energy in the cosmological setting allowing a clear interpretation of the first law of thermodynamics for de Sitter space. The derivation via horizon thermodynamics turned out to be remarkably simple providing insights also to spacetimes that are not to a good approximation de Sitter. The thermal argumentation implied that the fate of the Universe is in fact an asymptotically flat space instead of eternal de Sitter expansion.

The possible decay or evaporation of the de Sitter horizon seems like a prime candidate for explaining the unnaturally small amount of vacuum energy that is consistent with observations. Importantly, in our prescription for semi-classical gravity a gradual decrease of $H$ is recovered. Unfortunately, the predicted change is quite slow. For the Early Universe and in particular inflation the gradual decrease of $H$ from back reaction is much smaller than the slow-roll behaviour usually encountered in inflationary cosmology. Of course due to the multitude of various models of inflation, an evaporation mechanism could potentially provide a novel block for model building in at least some cases.

Perhaps the most profound implication of this work is that it suggests that potentially the eventual de Sitter evolution of the Universe as predicted by the current standard model of cosmology, the $\Lambda$CDM model, is not eternal. This indicates that at least some of the problems associated with the finite temperature and entropy of eternal de Sitter space and in particular the issues with Boltzmann Brains \cite{Carroll:2017gkl} could be ameliorated.

Of course all of the perhaps rather significant predictions from this work rest on the coarse graining prescription we have introduced in the semi-classical approach to gravity. Quite unavoidably, it results in an inherently observer-dependent approach due to the observer dependence of the de Sitter horizon. This is in accord with the statements of \cite{Gibbons:1977mu}, but from a fundamental point of view appears to result in rather profound conclusions such as Everett - Wheeler or many-worlds interpretation of quantum mechanics, as discussed in \cite{Gibbons:1977mu}. In a semi-classical approximation however no obvious inconsistencies seem to arise when one simply includes the additional step of coarse graining the quantum state with respect to the perceptions of a particular observer, although more work in this regard is required. 

Coarse graining over unobservable information gives rise to several natural features: it allows for the generation of entropy, the quantum-to-classical transition via decoherence and by definition leads to a result containing only the information an observer may interact with. When tracing over information beyond the event horizon of de Sitter space it also leads to an energy-momentum with a divergence on the horizon, which may signal a breakdown of the semi-classical approach but more investigation is needed. We end by emphasizing that in this work we have not presented a complete analysis of all physical implications of the prescription, which needs to be done in order to ultimately determine its viability. 

\acknowledgments{We thank Paul Anderson and Emil Mottola for illuminating discussions and Thanu Padmanabhan for valuable comments on the draft. This research has received funding from the European Research Council under the European Union's Horizon 2020 program (ERC Grant Agreement no. 648680) and the STFC grant ST/P000762/1.}

\appendix
\section{Details on mode normalization}
\label{sec:B}
Here we provide the details for the normalization in section \ref{sec:4}.

Starting with the mode defined in the unobservable patch of the expanding half of the de Sitter manifold (\ref{eq:ans2}) we first introduce an infinitesimal convenience factor in the arguments of the hypergeometric function
\ee{k\longrightarrow k - i 2\epsilon H\,,\label{shift}}
which gives \ea{&f^\text{\tiny \it  B}_{\ell k}(\ubar{r}_\text{\tiny \it  B})=D^\text{\tiny \it  B}_{\ell k}(H\ubar{r}_\text{\tiny \it  B})^\ell\big[(H\ubar{r}_\text{\tiny \it  B})^2-1\big]^{-\f{ik}{2H}}\times\nonumber \\ &\!\!{\,}_2F_1\bigg[\f{\ell}{2} -\f{ik}{2H} +\f{1}{2}-\epsilon,\f{\ell}{2} -\f{ik}{2H} +{1}-\epsilon;\ell +\f{3}{2};(H\ubar{r}_\text{\tiny \it  B})^2\bigg]\,.\label{eq:fb2}}
We can then use the standard relation
\ea{&{\,}_2F_1\big[a,b;c;z\big]=\frac{\Gamma[c]\Gamma[c-a-b]}{\Gamma[c-a]\Gamma[c-b]}\nonumber\\&\times\!{\,}_2F_1\big[a,b;a+b+1-c;1-z\big]+\frac{\Gamma[c]\Gamma[a+b-c]}{\Gamma[a]\Gamma[b]}\nonumber \\&\times(1-z)^{c-a-b}{\,}_2F_1\big[c-a,c-b;1+c-a-b;1-z\big]\,,\label{for}}
to write two important results
\ea{&\!{\,}_2F_1\bigg[\f{\ell}{2} -\f{ik}{2H} +\f{1}{2}-\epsilon,\f{\ell}{2} -\f{ik}{2H} +{1}-\epsilon;\ell +\f{3}{2};(H\ubar{r}_\text{\tiny \it  B})^2\bigg]\nonumber \\&\overset{H\ubar{r}_\text{\tiny \it  B}\rightarrow 1}{=}\f{\Gamma\big[\ell +\f{3}{2}\big]\Gamma\big[\f{ik}{H}
		\big]}{\Gamma\big[\f{1}{2}\big(\ell + \f{ik}{H} +1\big)
		\big]\Gamma\big[\f{1}{2}\big(\ell + \f{ik}{H} +2\big)
		\big]}+{\cal O}(\epsilon)\,,\label{res1}}
and
\ea{&\big[(H\ubar{r}_\text{\tiny \it  B})^2-1\big]\f{\partial}{\partial r}{\,}_2F_1\bigg[\f{\ell}{2} -\f{ik}{2H} +\f{1}{2}-\epsilon,\f{\ell}{2} -\f{ik}{2H} +{1}-\epsilon;\nonumber \\&\ell +\f{3}{2};(H\ubar{r}_\text{\tiny \it  B})^2\bigg]\overset{H\ubar{r}_\text{\tiny \it  B}\rightarrow 1}{=}0\,.\label{res2}}
The validity of (\ref{res1}) and (\ref{res2}) is only strictly true with the infinitesimal shift (\ref{shift}) introducing the factor \ee{\propto \big[(H\ubar{r}_\text{\tiny \it  B})^2-1\big]^{2\epsilon}\,,}
which suppresses the second term coming from (\ref{for}). With (\ref{res1}) and (\ref{res2}) one may verify the correct normalization of (\ref{eq:fb2}) by evaluating the inner product (\ref{eq:ippe}) at the time instant $H\ubar{r}_\text{\tiny \it  B}\rightarrow 1$. Equation (\ref{res1}) also allows one to easily show that close to the horizon with (\ref{eq:fb2}) the mode becomes a simple plane wave.

Introducing a similar infinitesimal factor in the hypergeometric function of the mode in the observable patch of the expanding half of the de Sitter manifold (\ref{eq:ans}) with $k\rightarrow k + i 2\epsilon H$ and using (\ref{res1}) we can verify that the horizon limit is again a plane wave with a prefactor coinciding with the horizon limit of (\ref{eq:fb2}), which shows that it is also correctly normalized.

Strictly speaking, by introducing the infinitesimal $\epsilon$ factor we are effectively solving a different equation of motion than when starting with $\epsilon=0$. 
However, at the end of a calculation once a physical observable has been derived we set $\epsilon\rightarrow0$ to obtain agreement with results obtained without the shift (\ref{shift}). For our purposes introducing the $\epsilon$ is useful as it simplifies some of the intermediate steps of the derivation.

We acknowledge fruitful discussions with the authors of \cite{Higuchi:2018tuk}.
\section{Energy density far from the horizon}
\label{sec:A}
In this appendix we calculate the energy momentum tensor in the static coordinates (\ref{eq:staA}) with the line element (\ref{eq:sta}) in a region close to the center of the Hubble sphere.
 
In order to calculate the expectation value of the energy density (again dropping the $A$ labels)
\ee{ \langle\hat{T}_{\,\ubar{\!0}\,\ubar{\!0}}\rangle=\langle\f{1}{2}\nabla_\rho \hat{\phi}\nabla^\rho \hat{\phi}+\nabla_{\,\ubar{\!0}}\hat{\phi}\nabla_{\,\ubar{\!0}} \hat{\phi} +\f{1}{6}\big[G_{\,\ubar{\!0}\,\ubar{\!0}}-\nabla_{\,\ubar{\!0}}\nabla_{\,\ubar{\!0}}-\Box\big] \hat{\phi}^2\rangle\label{eq:munu2}\,,}
it proves convenient to make use the equation of motion $-\Box\hat{\phi}=-2H^2\hat{\phi}$ and the relation \ee{\nabla_{\,\ubar{\!0}}\nabla_{\,\ubar{\!0}}\langle\hat{\phi}^2\rangle=\mathcal{O}(H\ubar{x})^2\,,\label{eq:simpp}} to arrive at the expression
\ea{ \langle\hat{T}_{\,\ubar{\!0}\,\ubar{\!0}}\rangle&=\langle\f{1}{6}\Big[
g^{\,\ubar{\!0}\,\ubar{\!0}}\big(\partial_{\,\ubar{\!0}}{\hat{\phi}}\big)^2+g^{\ubar{r}\ubar{r}}\big(\partial_{\ubar{r}}{\hat{\phi}}\big)^2+g^{\ubar{\theta}\ubar{\theta}}\big(\partial_{\ubar{\theta}}{\hat{\phi}}\big)^2\nonumber \\&+g^{\ubar{\varphi} \ubar{\varphi}}\big(\partial_{\ubar{\varphi}}{\hat{\phi}}\big)^2- H^2{\hat{\phi}^2}\Big]+\big(\partial_{\,\ubar{\!0}}{\hat{\phi}}\big)^2\rangle\label{eq:munu3}+\mathcal{O}(H\ubar{x})^2
\,.}
The simplifications (\ref{eq:simpp}) and (\ref{eq:munu3}) follow from the fact that to leading order in $\mathcal{O}(H\ubar{x})$ the radial contribution to solutions (\ref{eq:ans}) simplifies significantly and that the static line element coincides with Minkowski space. To this accuracy the only relevant modes are
 \ea{\psi_{0 0 k}&=D_{0k}Y^0_0 e^{-ik\ubar{t}}+\mathcal{O}(H\ubar{x})^2\\
 \psi_{m 1 k}&=D_{1k}H\ubar{r}Y^m_1 e^{-ik\ubar{t}}+\mathcal{O}(H\ubar{x})^2\,;\qquad m\in\{-1,0,1\}\,,}
 where from (\ref{eq:ncost}) we have
 \ee{|D_{0k}|^2=\f{k}{\pi}\,,\qquad |D_{1k}|^2=\f{H^2k+k^3}{9H^2\pi}\,,}
 and the spherical harmonics read
\ea{Y_{0}^{0}&={\f{1}{2}}\sqrt{\f{1}{\pi}}\,, \\
Y_{1}^{-1}&={\f{1}{2}}\sqrt{\f{3}{2\pi}} \, \sin\theta \, e^{-i\varphi}\,, \\
Y_{1}^{0}&={\f{1}{2}}\sqrt{\f{3}{\pi}}\, \cos\theta\,, \\
Y_{1}^{1}&={\f{-1}{2}}\sqrt{\f{3}{2\pi}}\, \sin\theta\, e^{i\varphi}\,.}
With the above, we can now evaluate the expression (\ref{eq:munu3}) piece-by-piece. The first is
\ea{\langle g^{\,\ubar{\!0}\,\ubar{\!0}}\big(\partial_{\,\ubar{\!0}}{\hat{\phi}}\big)^2\rangle &=-\langle\big(\partial_{\,\ubar{\!0}}{\hat{\phi}}\big)^2\rangle\nonumber \\&=-\Big\langle\bigg\{\int_0^\infty dk\Big[-ik\psi_{00 k}\hat{a}_{00k}+{\rm H.C.}\Big]\bigg\}^2\Big\rangle\nonumber\\
&=-\int_0^\infty \f{dk\,k^3}{4\pi^2}\Big[1+2\langle\hat{n}_{00k}\rangle\Big]\nonumber \\
&=-\int\f{d^3\mathbf{k}}{(2\pi)^3}\bigg[\f{k}{2}+\f{k}{e^{2\pi k/H}-1}\bigg]\,,\label{eq:ttt}}
where we have left the accuracy $\mathcal{O}(H\ubar{x})^2$ implicit and used 
\ea{\langle\hat{a}^{\dagger}_{\ell m k}\hat{a}^{\phantom{\dagger}}_{m'\ell' k'}\rangle&=\delta_{mm'}\delta_{\ell\ell'}\delta(k-k')\langle\hat{n}_{\ell m k}\rangle\nonumber \\&=\delta_{mm'}\delta_{\ell\ell'}\delta(k-k')\f{1}{e^{2\pi k/H}-1}\,,}
valid for the thermal density matrix (\ref{eq:desnm2}). 
We can continue in a similar fashion
\ea{\langle g^{\ubar{r}\ubar{r}}\big(\partial_{\ubar{r}}{\hat{\phi}}\big)^2\rangle&=\Big\langle\bigg\{\partial_{\ubar{r}}\!\!\sum_{m=-1}^1\int_0^\infty dk\Big[\psi_{m1 k}\hat{a}_{m1k}+{\rm H.C.}\Big]\bigg\}^2\Big\rangle\nonumber \\ 
&=\int_0^\infty dk\f{H^2k+k^3}{9\pi} 
\sum_{m=-1}^1 |Y^m_1|^2\nonumber \\&\times\Big[1+2\langle\hat{n}_{m1k}\rangle\Big]\nonumber \\
&=\f{1}{3}\int dk\f{H^2k+k^3}{4\pi^2} 
\Big[1+\f{2}{e^{2\pi k/H}-1}\Big]\,.}
Repeating these steps for the remaining contributions in (\ref{eq:munu3}) gives up to $\mathcal{O}(H\ubar{x})^2$
\ea{\langle g^{\ubar{r}\ubar{r}}\big(\partial_{\ubar{r}}{\hat{\phi}}\big)^2\rangle&=\langle g^{\ubar{\theta}\ubar{\theta}}\big(\partial_{\ubar{\theta}}{\hat{\phi}}\big)^2\rangle=\langle g^{\ubar{\varphi} \ubar{\varphi}}\big(\partial_{\ubar{\varphi}}{\hat{\phi}}\big)^2\rangle\nonumber \\&=\f{1}{3}\langle\big(\partial_{\,\ubar{\!0}}{\hat{\phi}}\big)^2\rangle+\f{H^2}{3}\langle{\hat{\phi}^2}\rangle\,,}
so the terms in the square brackets of (\ref{eq:munu3}) cancel so that (\ref{eq:munu3}) and (\ref{eq:ttt}) finally give
\ea{\langle\hat{T}_{\,\ubar{\!0}\,\ubar{\!0}}\rangle&=\langle\big(\partial_{\,\ubar{\!0}}{\hat{\phi}}\big)^2\rangle =\int\f{d^3\mathbf{k}}{(2\pi)^3}\bigg[\f{k}{2}+\f{k}{e^{2\pi k/H}-1}\bigg]\nonumber \\&+\mathcal{O}(H\ubar{x})^2\,,}
as written in (\ref{eq:t00}).

\end{document}